\documentclass[12pt]{iopart}
\usepackage{graphicx}
\usepackage{tikz}
\usepackage{verbatim}
\usepackage{array}
\usepackage[hang]{subfigure}
\usepackage{tabularx}
\usepackage{hyperref}

\newtheorem{theorem}{Theorem}[section]

\newtheorem{algorithm}[theorem]{Algorithm}

\def\complaint#1{}
\def\withcomplaints{
\newcounter{mycomplaints}
\def\complaint##1{\refstepcounter{mycomplaints}%
\ifhmode%
\unskip%
{\dimen1=\baselineskip \divide\dimen1 by 2 %
\raise\dimen1\llap{\tiny -\themycomplaints-}}\fi%
\marginpar{\tiny [\themycomplaints]: \textcolor{revision} {##1}}}%
}

\definecolor{adnan}{rgb}{0,0,1}
\definecolor{revision}{rgb}{0,1,0}

\makeatletter
\def\Ddots{\mathinner{\mkern1mu\raise\p@
\vbox{\kern7\p@\hbox{.}}\mkern2mu
\raise4\p@\hbox{.}\mkern2mu\raise7\p@\hbox{.}\mkern1mu}}
\makeatother

\def\picture #1 by #2 (#3){
   \vbox to #2{
     \hrule width #1 height 0pt depth 0pt
    \vfill
\special{picture #3} } }

\def\scaledpicture #1 by #2 (#3 scaled #4) (#5) {{
    \dimen0=#1 \dimen1=#2
   \divide\dimen0 by 1000 \multiply\dimen0 by #4
    \divide\dimen1 by 1000 \multiply\dimen1 by #4
\goodbreak
\midinsert\centerline{  \picture \dimen0 by \dimen1 (#3 scaled #4)}
{\eightbf
 {
  \noindent Figure~#3.}  {\eightrm #5\/} }\endinsert }}

\begin{document}

\title [Probing Protein Ensemble Rigidity and Hydrogen-Deuterium exchange] {Probing Protein Ensemble Rigidity and Hydrogen-Deuterium exchange}

\author{Adnan Sljoka$^1$\footnote{Supported in part under a grant from NSERC (Canada). Email: adnanslj@matstat.yorku.ca} and Derek Wilson$^2$\footnote{Supported by NSERC (Canada) Discovery Grant \# 341925. Email: dkwilson@yorku.ca}}

\address{$^1$Department of Mathematics and Statistics, York University,\\ 4700 Keele Street, Toronto, M3J 1P3,
 Canada}
\address{$^2$ Chemistry Department, York University,\\ 4700 Keele Street, Toronto, M3J 1P3,
Canada}



\begin{abstract}



Protein rigidity and flexibility can be analyzed accurately and
efficiently using the program FIRST. Previous studies using FIRST
were designed to analyze the rigidity and flexibility of proteins
using a single static (snapshot) structure. It is however well
known that proteins can undergo spontaneous sub-molecular
unfolding and refolding, or conformational dynamics, even under
conditions that strongly favour a well-defined native structure.
These (local) unfolding events result in a large number of conformers that differ from each other very slightly. In this context, proteins are better represented as a thermodynamic ensemble of `native-like' structures, and not just as a single static low-energy structure.


Working with this notion, we introduce a novel FIRST-based
approach for predicting rigidity/flexibility of the protein ensemble by
(i) averaging the hydrogen bonding strengths from the entire ensemble and (ii) by refining the
mathematical model of hydrogen bonds. 
Furthermore, we combine our FIRST-ensemble rigidity predictions
with the ensemble solvent accessibility data of the backbone
amides and propose a novel computational method which uses both
rigidity and solvent accessibility for predicting
hydrogen-deuterium exchange (HDX). To validate our predictions, we
report a novel site specific HDX experiment which characterizes the
native structural ensemble of Acylphosphatase from
hyperthermophile Sulfolobus solfataricus (Sso AcP).


The sub-structural conformational dynamics that is observed by HDX
data, is closely matched with the FIRST-ensemble
rigidity predictions, which could not be attained using
the traditional single `snapshot' rigidity analysis.
Moreover, the computational predictions of regions that are
protected from HDX and those that undergo exchange are in very
good agreement with the experimental HDX profile of Sso AcP.

\end{abstract}



\section{Introduction}

It has long been accepted that protein structural flexibility
and dynamics is as critical for protein function as its
3D-structure \cite{proteingeometrybook, nativestateconformations}. Accurate measurements of
flexibility and dynamics of proteins can help us interpret
the relationship between structure and function, and has significant biological implications in medicine and drug design \cite{goodallosteryreview, flexgraph, gohlkekuhn, residueflexibilitysolventaccessibility}. 
Being able to give fast predictions of flexible and rigid regions in the proteins and its dynamics is an important area of research in computational biology, particulary in high throughput studies.



Determining flexible and rigid regions in a protein and understanding
its motions is a complex task. The main difficulty is that conformational fluctuations are rapid, transient and result in structures that are spectroscopically indistinguishable from the ground-state. 
A wide
range of experimental data (NMR techniques such as order parameter measurements, relaxation dispersion, hydrogen/deuterium exchange data, etc.) can provide some limited insights \cite{experimentalnmrflexibility,
hdexchangehvidt, experimentalflexibility}. Computational methods have also facilitated enormous strides in this area \cite{nmarigidityapplicationbook, morphserver, flexgraph, nma2, nussinovhinges}. Molecular dynamics (MD) simulations is traditionally used to probe protein mobility and flexibility, but its main downside is that it takes a prohibitive amount of computational power to investigate the functionally relevant micro- and millisecond time-scales.


Given the rapid growth in the number of entries in the protein
data bank and the size of the solved protein structures, combined
with the severe limitation of computational power and resources
needed to study protein flexibility and dynamics with traditional
methods, there is a tremendous need to develop faster
computational methods. One such computationally fast new method is
FIRST \cite{first, flexgraph} (Floppy Inclusion and Rigid
Substructure Topography) or its earlier version PROFLEX \cite{proflex} along with its extensions, such as FRODA \cite{froda}.

FIRST (see Section \ref{sec:first}) is based on a well established mathematical theory of Rigidity Theory
\cite{whiteleyrigidbook, countingout} and concepts in solid state physics
\cite{firstglasses}. 
Computationally, FIRST uses the combinatorial pebble game algorithm \cite{LeeStreinu, adnanthesis, countingout}, which performs the constraint
counting in the molecular multigraph (constraint network) in order
to match the degrees of freedom with various biochemical
constraints, and outputs all the rigid and flexible regions in the
protein. 
While considerable computational power is needed to study protein flexibility with MD simulations, FIRST can predict the rigid clusters and flexible connections (known as the \emph{rigid cluster decomposition}) in a matter of seconds. 
Numerous studies have thoroughly demonstrated that FIRST gives accurate predictions of flexibility and rigidity in proteins
\cite{gohlkekuhn, cores, rigiditylost, flexgraph} and RNA
\cite{rnafirst}. 

\subsection{Extending FIRST to Protein ensembles}

In the `new view' of protein folding and energy landscapes, the native state is represented as the minimum in a narrow, symmetric free energy well \cite{dill}. Even under native conditions, proteins occupying this well can undergo local conformational fluctuations that occur on a wide range of time-scales, from microseconds to hours, sampling a distribution of native like conformational substates, referred to as the `native-state ensemble' \cite{Englander, nativestateconformations, landscape, hdexchangehvidt, goodallosteryreview, manussinov, conformationalensembles, hdensemblesadqi}. 
The native-state ensemble is important to aspects of protein function including stability, cooperativity and catalysis \cite{eisenmesser, khrishna, labeikovsky, manussinov, wolfwatz}. There is therefore a strong incentive to characterize the native structural ensemble with a view to isolating those conformers that are most associated with biological activity.

One major source of experimental data that can provide structural
insight on the native state conformational ensemble is available
in the protein data bank (PDB) \cite{pdb}. NMR structures in the PDB are consistently expressed as
`ensembles' of structures, where the entire ensemble 
represents a possible solution set to the NMR structure
determination problem \cite{nmr3}. Some promising work in the area of ensemble refinement suggests that structures solved by x-ray crystallography, where discrete conformations of the atomic coordinates can be identified, would be better represented as a
set of multiple conformers \cite{ensemblecrystal2, ensemblecrystal1}.   Unfortunately, to date, only a small number of X-ray crystal structures with several conformers are available in the PDB \cite{ensemblecrystal1}.

One clear limitation of the current FIRST method (and equivalent implementations) is that it is primarily designed to perform the flexibility analysis using a single structure (snapshot) of a protein. When the protein structure is represented with a collection of conformers, particularly the NMR solved structures, previous FIRST studies would perform the rigidity/flexibility analysis only on a selected single structure (i.e. typically using the `most-representative' structure-model defined by the authors of the NMR structure), completely neglecting the information encoded in the other structures of the ensemble. In fact, there has been no distinction in the FIRST analysis of X-ray structures with single snapshots and those from NMR solved structures.

By selecting only a single NMR structure, not only is the crucial
ensemble information omitted, but the development of the
constraint network purely from the single structure can make the
rigidity/flexibility analysis more sensitive to the quality of the
structure (snapshot) selected \cite{flexgraph}. It is well known that vary small structural variations in the constraint network (i.e. breaking of a few hydrogen bonds and in some extreme cases a single hydrogen bond) can have a significant
effect on the rigid cluster decomposition, breaking a single rigid
cluster into numerous smaller rigid regions
\cite{thermostabilitygohlke, Wells}. 
It is then natural to hypothesize that analyzing multiple snapshots should alleviate these sensitivities and
deficiencies of the FIRST analysis, in particular the dependence of rigidity predictions on the selection of modelled non-covalent interactions, especially hydrogen bonds. Clearly, better ways of predicting rigidity of conformational ensemble than the current FIRST model are desirable. 



A major objective of this paper is to enhance the predictive power of FIRST for dynamics by conducting the analysis on the ensemble.
We will achieve ensemble-based prediction from FIRST by averaging the hydrogen bonding interactions over all the individuals structures (i.e. NMR models) of the ensemble, and by a refinement of the mathematical model of hydrogen bonds, to give a single ensemble FIRST prediction (see Materials and Methods section). We will illustrate that the FIRST-ensemble predictions can overcome some of the limitations of the traditional single (static) snapshot analysis, and should provide us with a more sensible and improved
picture of rigidity/flexibility.

\begin{figure}[h!]
\centering
   \subfigure[] { \includegraphics[width=.4\textwidth]{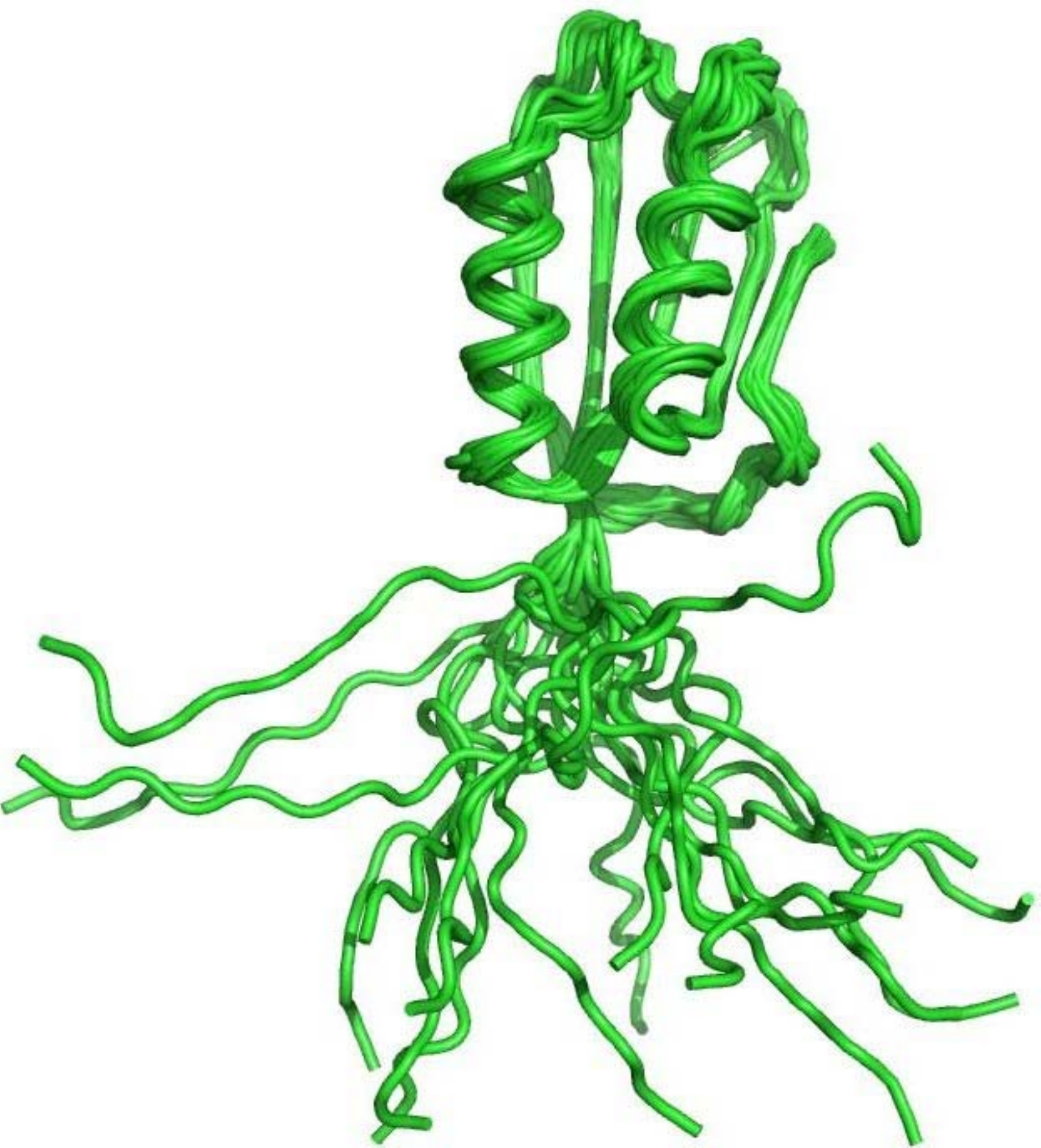}} \quad
   \subfigure[] { \includegraphics[width=.4\textwidth]{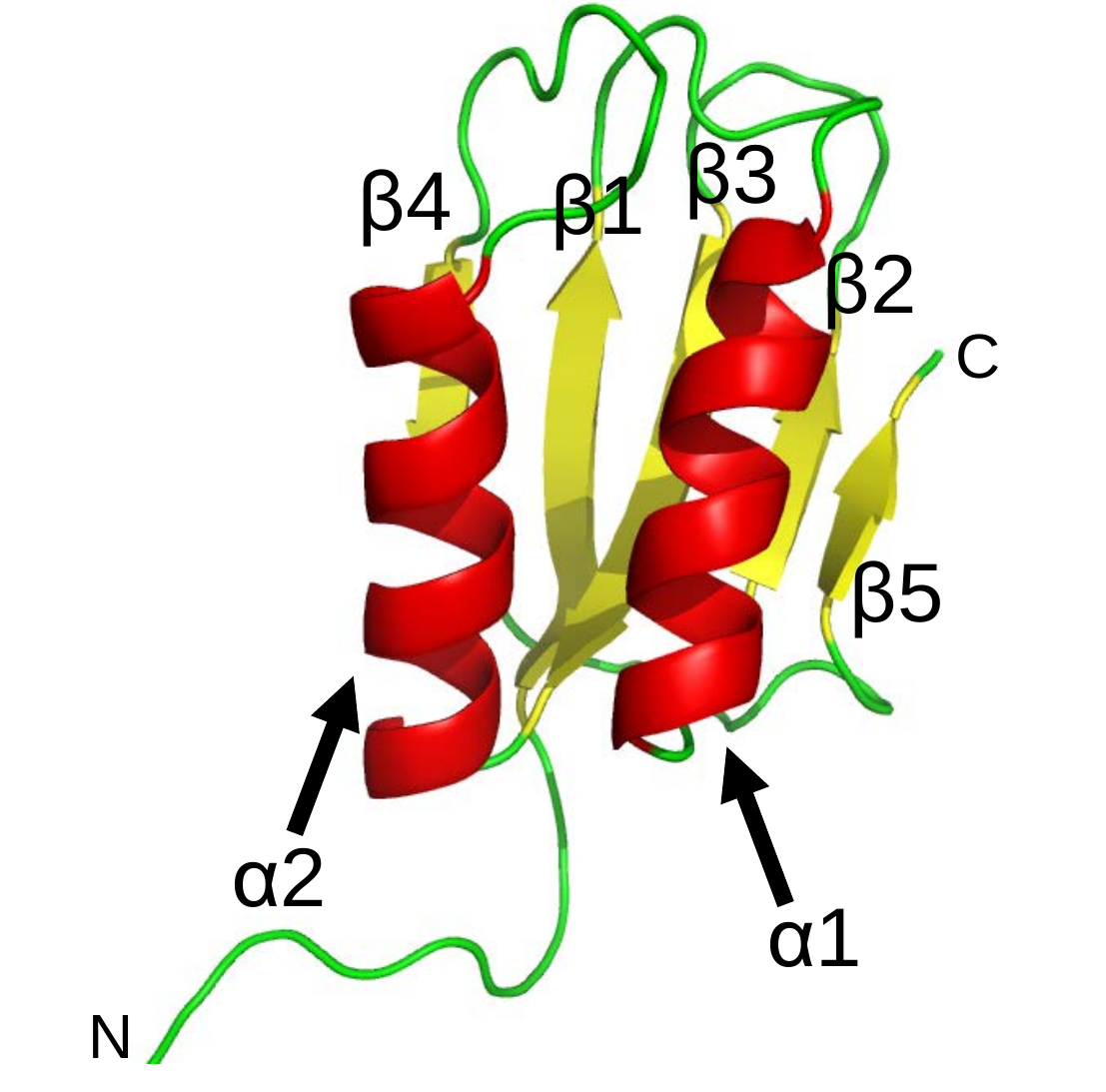}} \quad
    \caption{NMR derived 3-dimensional structure of Acylphosphatase
from hyperthermophile Sulfolobus solfataricus (Sso AcP) (pdb id:
1y9o). (a) The entire
structural ensemble (20 models) is shown in ribbon representation
(images generated with Pymol \cite{pymol}) (b) the first model (i.e. most representative structure) is
shown in the cartoon representation and coloured and labeled according to its secondary structure.}
    \label{fig:1y9o}
    \end{figure}

To test our method we will apply it to the native structural NMR
ensemble of Acylphosphatase from hyperthermophile Sulfolobus
solfataricus (Sso AcP) (figure \ref{fig:1y9o}). Sso
AcP is a 101-residue protein, which belongs to the
acylophosphatase-like structural family \cite{1y9opaper1}.The hyperthermophile nature of Sso AcP also offers a unique opportunity
to apply the FIRST analysis to an enzyme that is expected
to be non-functional at room temperature due to rigidity. We will compare the traditional FIRST single snapshot rigidity  prediction with our modified FIRST prediction over the entire ensemble.  The validity of these predictions is then tested by comparison with an experimentally derived picture of the ensemble acquired via NMR-based hydrogen/deuterium exchange (HDX) measurements. Our FIRST-ensemble approach is quite general and can equally be applied to an ensemble of snapshots generated by other techniques, for instance conformers extracted from coarse-grain MD simulations
\cite{gohlkekuhn, flickering}.





\subsection{Combined Rigidity and Solvent Accessibility Analysis is a Computationally Facile Predictor of Hydrogen/Deuterium Exchange}



HDX is a powerful experimental technique as it can provide us with
direct information about protein's dynamics and structural stability \cite{Englander, englanderhd}. It is particulary useful as it is
sensitive to the entire structural ensemble, even to the most
rarely sampled high energy conformers. At physiological pH, HDX is a process in which amide protons on the polypeptide `backbone' and `exchangeable' protons on some side chains undergo base-catalyzed exchange with solvent \cite{englanderold}.
Experimental measurements of HDX are typically confined to the
backbone amide protons, which are involved in hydrogen bonding for
the maintenance of secondary structure. Since the hydrogen bond must first `break' in order for exchange to occur, HDX can be used to provide a semi-quantitative measurement of local thermodynamic stability in secondary structures. HDX can also occur when the amide hydrogens
become exposed (accessible) to the solvent \cite{englanderhd}.
Thus, the observed rate of backbone amide exchange is a function
of hydrogen bond strength and solvent accessiblity
\cite{englanderold, woodwardhd}.

Using NMR techniques it is possible to monitor backbone amide
proton signals as a function of time after exposure to deuterated
water (D$_{2}$O) (deuterium does not produce a signal in
conventional Heteronuclear Single Quantumn Coherence (HSQC)
experiments), and assign ensemble exchange rates to individual residues
(backbone amide protons). In our study, an experimental residue-specific HDX-profile was obtained for Sso Acp, and residues were classified into a range of categories, from very slow exchangers to very fast exchangers. Residues that do not exchange and residues that are very slow exchangers are either buried away from the solvent or they are found in very stable and rigid regions of a protein with strong hydrogen bonding interactions.

Although experimental methods for measuring HDX are well established (using NMR and Mass Spectrometry
technologies), they can be very costly and time
demanding. Only a few computational methods
have been suggested \cite{predictionhdxsequence, dxcorex}. Devising new fast algorithms that
can give rapid predictions of the regions in the protein that are
protected from HDX and regions that are not protected would be very valuable. For instance,
predictions of HDX (together with sequence homology) is used
to predict $3$-dimensional structures of unknown proteins
\cite{dxcorex}. By avoiding the bottlenecks associated with the slow experimental techniques, fast computational HDX prediction algorithms can also enable high throughput analysis.

As HDX depends on both structural stability and
solvent accessibility, we hypothesize that combining rigidity predictions with solvent accessibility data should provide us direct insight into
regions that are most likely protected from HDX and regions that are not protected from HDX. More specifically, if a region of a protein is rigid
(particulary over the ensemble) it will have a sufficient number
of constraints (a rich hydrogen bonding network) to prevent it
from undergoing HDX. Similarly, regions whose backbone
amides (NH) are buried (inaccessible) from the solvent, should not
be good exchangers. On the other hand, backbone amides in flexible and solvent accessible regions should be among the fast exchangers. 

In order to test this hypothesis, and as the main goal of the paper, we will introduce a new
computational method for predicting HDX, which combines our FIRST-ensemble rigidity predictions with computationally generated ensemble solvent
accessibility data (see materials and methods) of the backbone amide (NH).

In summary, the following two new contributions are based on this
work:

$\bullet$ development of an ensemble-based FIRST method for predicting protein flexibility/rigidity

$\bullet$ an introduction of a novel computational method which
combines FIRST-ensemble rigidity predictions and ensemble solvent
accessibility data as a predictor of HDX
(an ensemble measurement);

We will apply these techniques on the case study protein Sso AcP, and
the predictions will be compared with our experimentally obtained HDX
data. To our best knowledge, this is first analysis which uses both rigidity and solvent accessibility for predicting HDX.

\section{Materials and Methods}

\subsection{FIRST}
\label{sec:first}

Starting with the 3-dimensional snapshot of a protein structure (PDB file), FIRST generates a constraint multigraph (a graph which allows multiple edges between vertices) \cite{firstcapsid, countingout}, where the molecule is viewed as a body-hinge engineered structure of fixed units (atoms or bodies with their bond angles as rigid
units, bonds as potential hinges) plus other molecular constraints extracted from the local geometry. In the
constraint multigraph, vertices represent atoms and edges represent the distance constraints corresponding to covalent bonds, hydrogen bonds and hydrophobic interactions.

The
strength of each potential hydrogen bond is calculated based on
its local donor atom--hydrogen-acceptor atom (angular and distance) geometry (see
\cite{cores, flexgraph, Wells} for details). Hydrophobic contacts or tethers are modelled as any close contacts between pairs of carbon and/or sulfur atoms \cite{flexgraph}. 
Once all the constraints are considered and the user has selected
a hydrogen bond energy cutoff
value, the pebble game algorithm rapidly decomposes a multigraph (protein) into rigid clusters and flexible
regions \cite{cores, flexgraph, Wells}.
In every rigid cluster, all bonds will be non-rotatable and all its atoms can only move as a single rigid body. 
On the other hand, flexible regions lack sufficient number of constraints to further restrict their internal motions.
A protein will normally be composed of several large rigid clusters,
which are connected by flexible regions. For further details consult the FIRST user manual and references \cite{gohlkekuhn, Wells, cores, firstcapsid, flexgraph, rigiditylost}.

\subsection{Comments on Hydrogen Bonds}

The output of FIRST is almost entirely dependant on the set of modelled hydrogen bond constraints. Revisiting important features of hydrogen bonds and understanding the limitations of the current hydrogen bond mathematical rigidity model will help us consider refinements and facilitate a meaningful extension of FIRST predictions to structural ensembles.

The hydrogen bond energy cutoff will distinguish between weak and strong hydrogen bonds. However, no further quantitative distinction is made; any two hydrogen bonds that meet the threshold are included as constraints and remove the same number of degrees of freedom \cite{flexgraph, Wells}. In fact, in terms of the mathematical model of rigidity, the hydrogen bonds that pass the cutoff are modelled equivalently to the covalent bonds; for each hydrogen bond, five bars (edges) are placed between the hydrogen and the acceptor atom.

There are further uncertainties which are unique to hydrogen bonds that need to be considered. Molecular dynamics simulations have confirmed that many hydrogen bonds have short life times and undergo `flickering' on the order of tens of picoseconds to a few nanoseconds \cite{flickering, flickering2}.  This is particularly true in the flexible and dynamic regions of the protein, where hydrogen bonds spontaneously break and reform to adjust to the  conformational changes of atoms. Every conformation that the protein samples under native conditions has a particular set of hydrogen bonds, which can be significantly different in another conformation due to the local changes in hydrogen bonding geometry. The authors of \cite{Wells} report that small variations in the donor-acceptor distance can lead to substantial changes in the hydrogen bond energy strengths. Consequently, FIRST runs on the snapshots generated with molecular dynamics simulation \cite{flickering} have significant differences in the rigid cluster decompositions. To address the `flickering' (time-dependant) nature of hydrogen bonds and the fluctuations in hydrogen bond strengths (i.e. geometry), Mamonova et al. \cite{flickering} have incorporated the lifetimes of hydrogen bonds (how often is each hydrogen bond present over the MD simulation) into FIRST analysis. This approach gave an improved FIRST prediction which better matched the experimental evidence and molecular dynamics simulations. 

Whenever we are given a collection of snapshots (i.e. NMR ensemble), in order to obtain more accurate and realistic FIRST predictions on the ensemble it will be crucial to consider a refinement of the current hydrogen bond model. A natural first step in this refinement is to obtain the average hydrogen bond strength for each hydrogen bond over all NMR conformers. Only hydrogen bonds that have strong average strengths (energy) over the entire ensemble will be included in the constraint multigraph (see Algorithm \ref{alg:firstensemble} for details). Furthermore, we will also modify the rigidity model of hydrogen bonds by considering the persistence of hydrogen bonds. Instead of always using 5 bars between the acceptor and donor atoms, we will allow the number of bars (edges) to vary between 1 and 5 (see figures \ref{fig:hbondsbarsnew} and \ref{fig:barassignment1}). By varying the number of bars, we can adjust the number of DOF that the hydrogen bond should remove based on its persistence over the ensemble. If the hydrogen bond does not persist over all snapshots, it will be subject to a penalty. In other words, if the average energy strength of a particular hydrogen bond is sufficiently high over the entire ensemble, yet it is not present (or very weak) in several snapshots, the hydrogen bond will still be included as a constraint. However, given that there will be a bar(s) penalty for this kind of hydrogen bond, a constraint with say 2 or 3 bars will remove less DOF from the overall system than the traditional 5 bar constraint. This permits us to gradually weaken those bonds that may not persist in all the snapshots (models).

This type of revised modelling of hydrogen bonds via averaging the strengths over an ensemble and varying the number of bars should give us a more tuned and refined representation of hydrogen bonds. It also facilitates the move away from the simple `on/off' type of modelling of hydrogen bonding constraints that has been applied in previous FIRST studies.

\subsection{First-ensemble procedure}

We now describe the procedure that gives a single FIRST prediction
(rigid cluster decomposition) on the entire ensemble from an NMR
protein structure. We call such predictions \emph{FIRST-ensemble}.
Typical NMR file will contain 20 models that best fit the NMR
data, so we take m = 20 below.

All the rigidity/flexibility runs on an NMR solved structure of
Sso AcP (pdb id: 1y9o) are performed using a standard FIRST
software implementation. In all the FIRST predictions on both
the individual NMR snapshots and on the FIRST-ensemble
predictions, we have set the hydrogen bond energy cutoff to $-$1.0 kcal/mol, that is we include only those
hydrogen bonds whose energy is less than (more favourable than)
$-$1.0 kcal/mol.

\begin{algorithm}
-- \textbf{FIRST-ensemble procedure:}

\textbf{Input}: NMR PDB file (or other source of protein ensembles).

\textbf{Output}: single FIRST rigidity/flexibility ensemble
prediction (i.e. rigid cluster decomposition).

\begin{itemize}
\item[(1.)] Run FIRST on every NMR model m (m = 1..20) using a
hydrogen bond energy cutoff of 0 kcal/mol and obtain the energy
strength E$_{i}$$^{m}$ for every hydrogen bond i.
\item[(2.)] Calculate the average energy strength E$_{AVG}$$_{i}$
for each hydrogen bond i over all 20 models:

E$_{AVG}$$_{i}$ $=$ $\frac{\sum_{m=0}^{20}E_{i}^{m}}{20}$
\item[(3.)] Use the following criteria to assign the bar (edge)
penalty between hydrogen and acceptor atoms for each hydrogen bond
i:
\begin{itemize}
\item[(a)] Define hydrogen bond i as being \emph{present} in a
specific model (snapshot) if its energy strength is less than
(more favourable than) $-$0.5 kcal/mol.
Denote the total number of
times hydrogen bond i is present out of 20 models as N$_{i}$.
\item[(b)] Assign the number of bars (edges) B$_{i}$
(\emph{correction factor}) for every hydrogen bond i, using the
following rule:
B$_{i}$ $=$ $\lceil$$\frac{N_{i}}{20}$$\times$ 5$\rceil$ (where
$\lceil$$x$$\rceil$ is a ceiling function, see figure
\ref{fig:barassignment1})
\end{itemize}
\item[(4.)] Select the first NMR model (any model could be
selected - see supplementary information and discussion below) and run FIRST
\end{itemize}

\label{alg:firstensemble}
\end{algorithm}

\begin{figure}[h!]
\centering
   \subfigure[] { \includegraphics[width=0.35\textwidth]{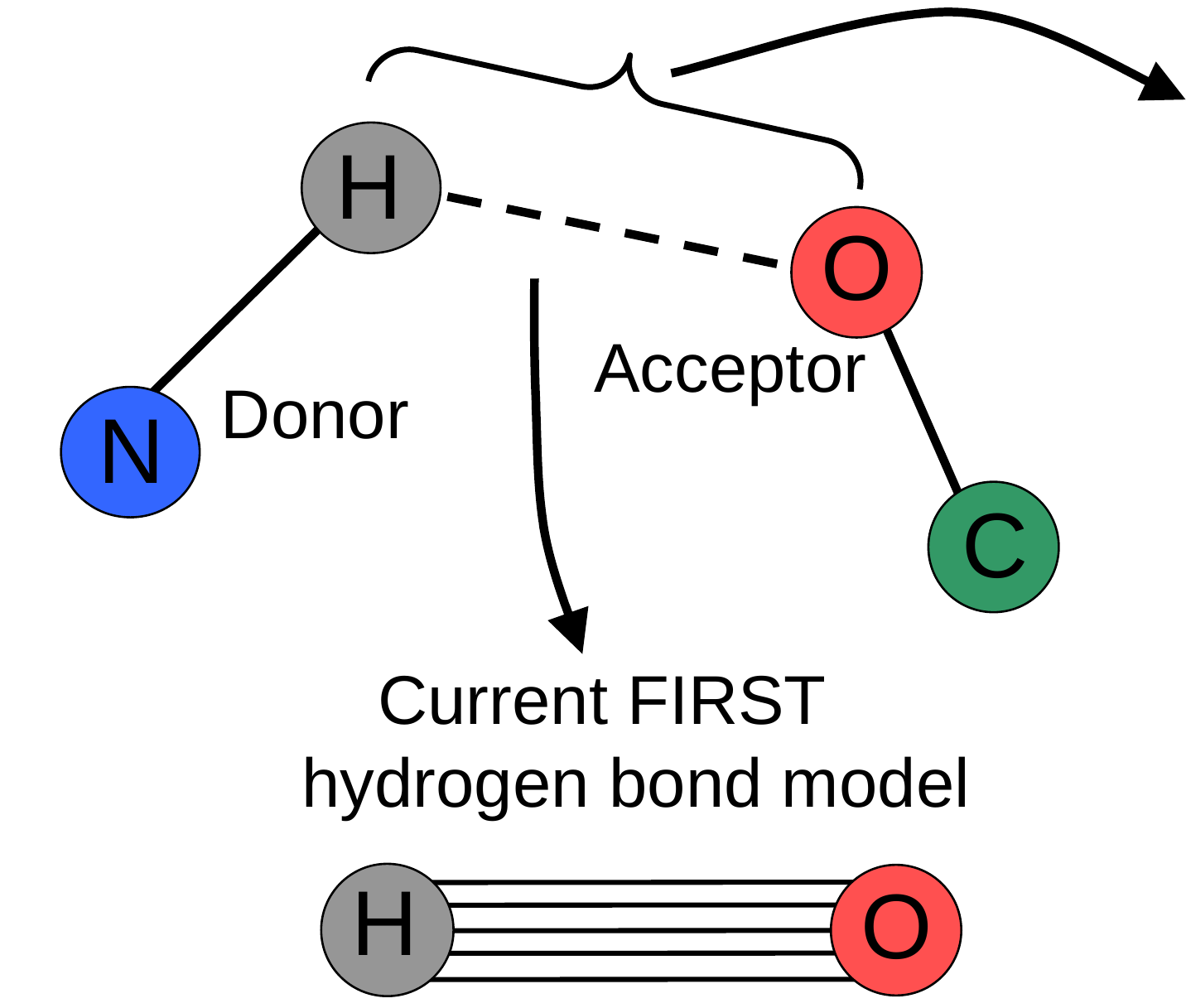}} \quad
   \subfigure[] { \includegraphics[width=0.28\textwidth]{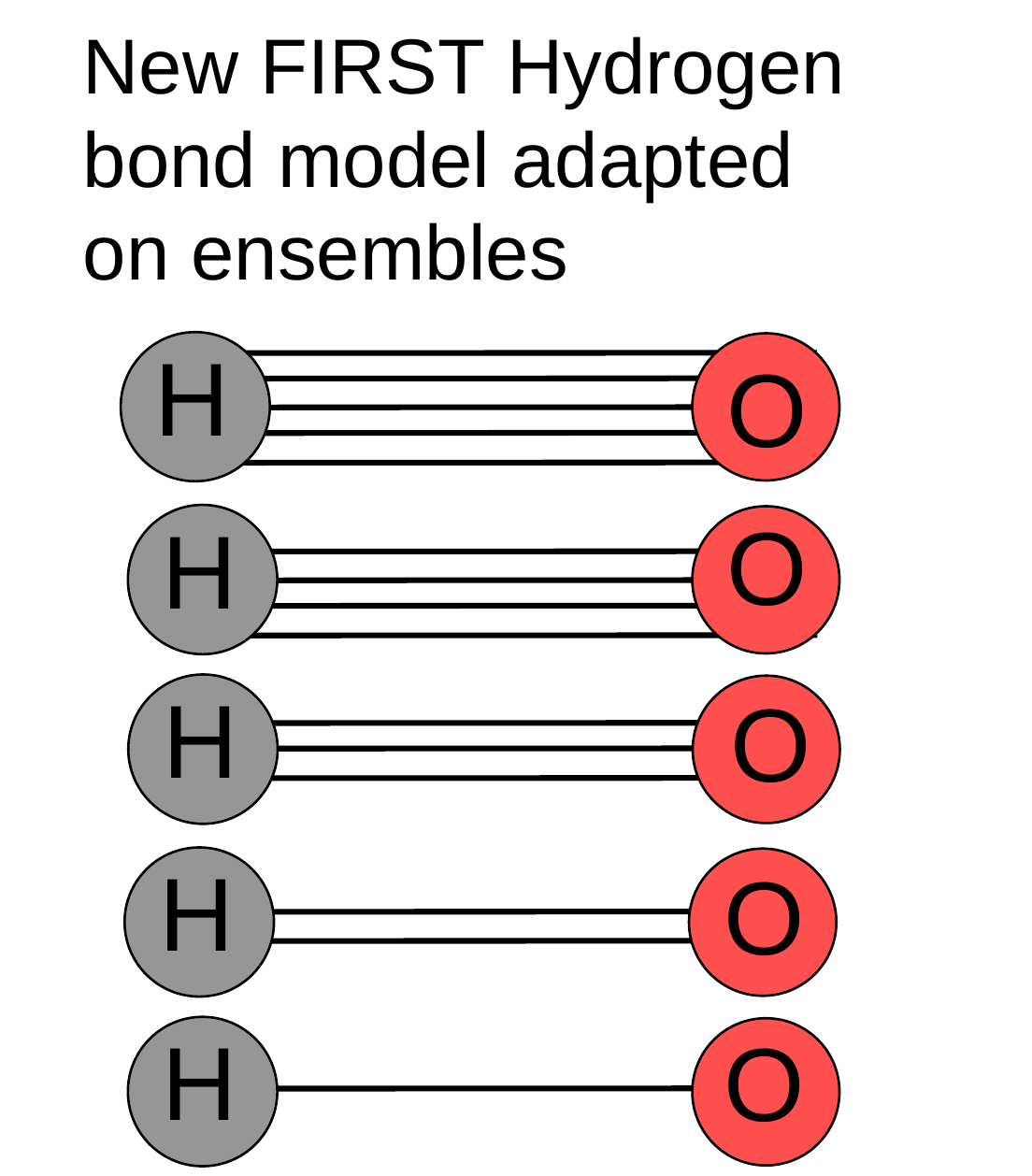}} \quad
    \caption{In FIRST hydrogen bonds are modelled with 5 bars (edges, constraints) between hydrogen and acceptor atoms (a). In the FIRST-ensemble predictions we allow the number of bars to vary between 1 and 5, depending on the persistence of a hydrogen bond over all structures in the ensemble (b).}
    \label{fig:hbondsbarsnew}
    \end{figure}

\begin{figure}[h!]
\centering
{\includegraphics[width=.8\textwidth]{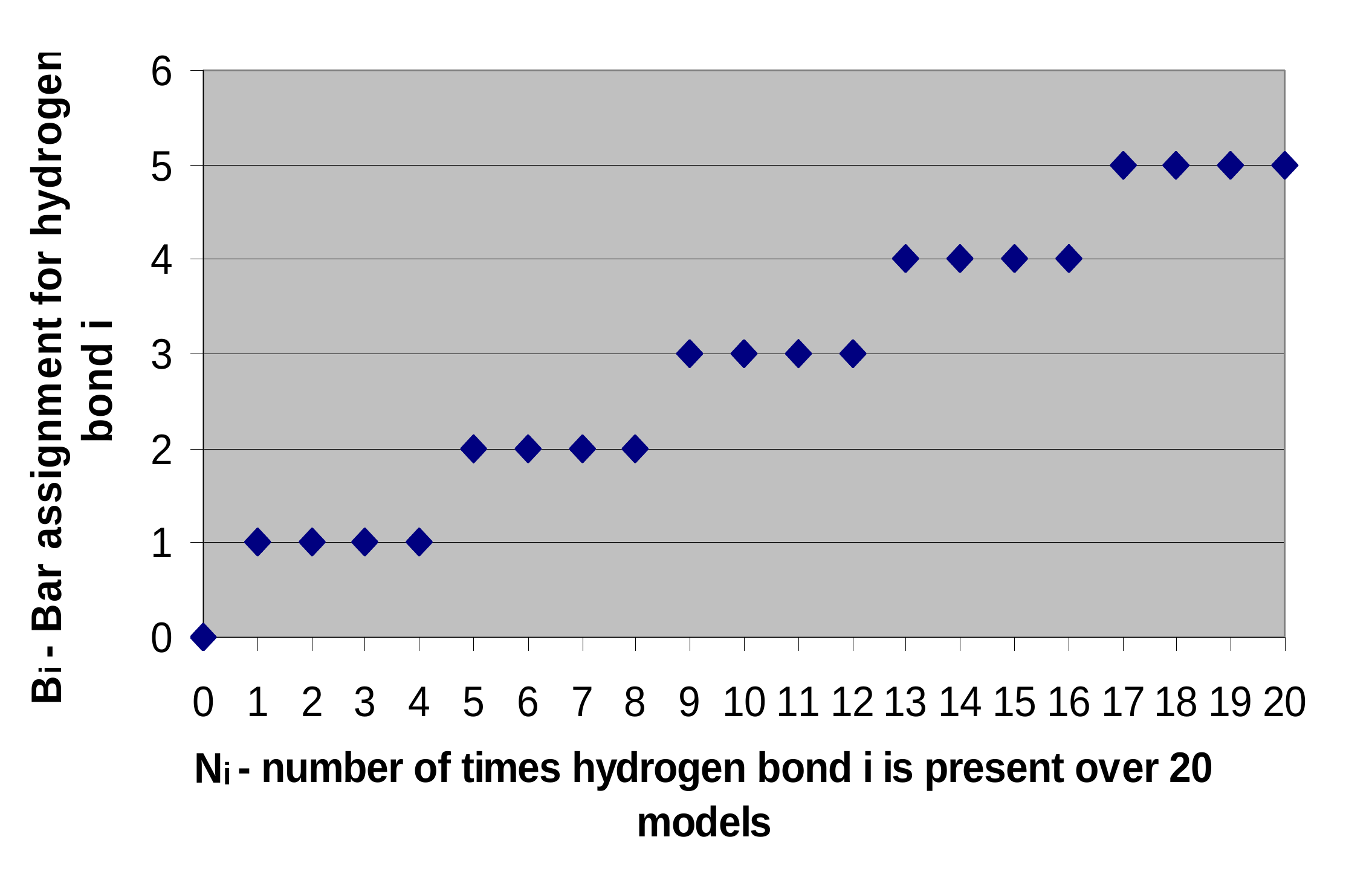}} \quad
    \caption{The number of bars (edges) B$_{i}$ that are assigned for hydrogen bond $i$ depends on how
    often the bond is present in the ensemble.}
    \label{fig:barassignment1}
    \end{figure}

In step 1. we run FIRST with energy cutoff of 0 kcal/mol
as this cutoff would consider all weak and strong hydrogen bonds. In the FIRST-ensemble approach, we still use an energy cutoff to
distinguish weak from strong hydrogen bonds (averaged over the
ensemble), with the added feature that the correction factor
B$_{i}$ allows us to incorporate the measure of the persistence of
a hydrogen bond over the ensemble. Consider for instance a hydrogen bond i that is present in 10 out of 20 models, with
relatively strong individual energies E$_{i}$$^{m}$ over the 10 models. The average
E$_{AVG}$$_{i}$ strength may still be relatively (negatively)
high, such that it passes the energy cutoff (in our study -1.0 kcal/mol). In this case this bond would be included as a constraint in the rigidity analysis. The usefulness of
correction factor B$_{i}$ can be seen as it takes into account the
absence (weak energy, bad geometry) of this hydrogen bond in the
remaining 10 models, giving it a 2 bar penalty, and an assignment
of 3 bars (see figures \ref{fig:hbondsbarsnew} and
\ref{fig:barassignment1}). 






\subsection{Solvent Accessibility Calculations}
\label{sec:accessibility}



The solvent `accessible surface area' (ASA) is the measure to which atoms on the protein surface are able to make contact with water. ASA is defined as the surface area traced out
by the center of a water sphere (radius of about 1.4
angstroms) as it is rolled over the surface of the protein \cite{chothia, whatif1}. It is a standard
practice to convert the ASA to its normalized form -- `relative
solvent accessibility' (RSA), which is the ratio of ASA in the current folded conformation to the ASA in the extended-unfolded conformation taken in the Gly-X-Gly tripeptide state \cite{rostaccessibility, whatif1}. 
Given that the experimental HDX on Sso AcP measures exclusively the HDX rates of backbone amide protons, we are only interested in obtaining
solvent accessibility of the backbone amides (NH). We will use the WHATIF \cite{whatif1} procedure to determine
the RSA for all backbone nitrogens of Sso AcP in all the 20 models of an NMR ensemble.
We then calculate the average (over the 20 NMR models) RSA for every backbone nitrogen.

We are primarily interested in finding the set of residues whose
(ensemble averaged) backbone amides (NH) are completely or almost
completely buried from the solvent, as these residues should not
be undergoing exchange. 
The reported RSA cutoffs to distinguish buried states in the literature have varied greatly, with values in the
0\% to 15\% range \cite{naccess, solventaccessibilitysecondary, accessibilityvarious, rostaccessibility}.
We define any residue as \emph{almost completely
buried} (or \emph{solvent inaccessible}) if its backbone nitrogen
RSA value (averaged over an entire ensemble) is less than 4 \%,
otherwise it is \emph{exposed} (or \emph{solvent
accessible}). This strong cutoff allows us to select those
residues whose backbone amide (NH) are closest to being completely
buried. A smaller accessibility cutoff was not chosen due to potential numerical rounding
approximations, and to eliminate the situations where a
slight increase in average accessibility is a result of
higher accessibility in one or two models.

When visualizing accessibility on a protein's 3D structure, almost
completely buried residues will be coloured blue. We further classify the exposed residues in the following categories (whose thresholds are somewhat arbitrarily selected but close to
\cite{rostaccessibility} for example): if the backbone nitrogen
has accessibility which is between 4 \% and 10 \% we say that the
corresponding residue is \emph{somewhat exposed} (yellow), between
11 and 30 \% \emph{mostly exposed} (orange), and greater than 30
\% \emph{almost completely exposed} (red) (see figure
\ref{fig:accessibility4colours}). The results on Sso AcP are not sensitive to these threshold
cutoffs and these remaining categories are solely assigned for easier
visualization and comparison.


\subsection{Combining FIRST-ensemble and Ensemble Solvent Accessibility as a Predictor of
HD-exchange} \label{sec:combining}

Both rigidity/flexibility and solvent accessibility hold valuable
information that can assist in predicting HDX. However, rigidity/flexibility or accessibility on its own provides only limited information. Consider a scenario where a given region of interest in the protein is flexible but its amide
protons are completely buried away from the solvent, which would make this
region a slow HD exchanger. If we did not have the solvent
accessibility data, sole rigidity/flexibility predictions
would not tell us about the lack of HDX in this
region. In contrast, consider a rigid region which is accessible to the
solvent. Using only solvent accessibility in this case is
insufficient to probe HDX, since in a typical rigid region of a
protein there is a significant number of strong hydrogen bonding
interactions that will not transiently `break' to allow HDX.

To probe our initial hypothesis, we now outline a procedure that combines the FIRST-ensemble rigidity predictions and solvent accessibility ensemble data, as a measure of HDX (an ensemble measurement).

\begin{algorithm}
-- \textbf{Rigidity and Accessibility as a prediction of HDX}

\textbf{Input}: NMR PDB file (or other source of protein ensembles).

\textbf{Output}: Prediction of regions which are most likely protected from HDX (no exchange or slow exchangers), and
regions which are likely to undergo exchange.

\begin{itemize}

\item[(1.)] Run the FIRST-ensemble procedure (Algorithm
\ref{alg:firstensemble}) to obtain the rigid cluster decomposition
for the ensemble.
\item[(2.)] Consider rigid clusters from (1.) and combine into a
single coloured \emph{rigidity-region}. Display this combined
rigidity-region with a unique colour on the protein's
3-dimensional structure, and choose a different colour for the
remaining flexible regions. (We will colour the combined
rigidity-region with the same colour as the largest rigid cluster,
which is blue by default, and intervening flexible regions with
gray.)

\item[(3.)] Superimpose the average solvent accessibility regions
(preferably using the colouring convention given in Section
\ref{sec:accessibility}) on the remaining flexible regions
(excluding $\alpha$-helices and prolines) of the protein from step
2. `Almost-completely buried' (inaccessible) regions should be
displayed with the same colour as the combined-rigidity region
(blue) in step (2.)

\end{itemize}

\textbf{The combined rigidity-region and almost-completely buried
(inaccessible) region (both coloured with blue) correspond to the
regions that we predict will most likely be protected from HDX
(i.e. slowest exchangers). The remaining regions: flexible (gray),
somewhat exposed (yellow), mostly exposed (orange), and almost
completely exposed (red) are the regions most likely not protected
from undergoing HDX (i.e. fastest exchangers).}
\label{alg:rigidityaccessibilitycombined}
\end{algorithm}

Due to the amphipathic
character of $\alpha$-helices, solvent accessibility is excluded for
$\alpha$-helices \cite{solventaccessibilitysecondary}. Solvent accessibility is not used for prolines as HDX is not assigned to prolines; amino acid
proline has no backbone amide hydrogen.

\subsection{Native State Hydrogen-Deuterium Exchange Experiment on Sso AcP}

\textbf{Materials and Protein Preparation and Purification.} Mono-
and di-basic sodium phosphate was purchased from Sigma (St. Louis,
MO). Ultrapure water was generated in-house on a Millipore
Advantage system. Uniformly $^{15}$N labeled protein was produced
from BL21(DE3) E. coli transformed using a pGEX-2T plasmid as
described previously \cite{hdx1}. Briefly, transformed cells were
grown for 8hrs at 38$^{o}$C in M9 media with
$^{15}$NH$_{4}$Cl$_{2}$ (Cambridge Isotope Laboratories, Andover,
MA) as the sole Nitrogen source. Overexpression of Sso AcP was
induced by IPTG when cultures reached OD $\sim$ 0.6 (typically
around 5 hrs). Cell lysis was by sonication. Sso AcP-GST was
purified from the supernatant on a glutathione column with
cleavage of the fusion protein by overnight digestion with
Thrombin.

\textbf{NMR Spectroscopy.} $^{1}$H-$^{15}$N HSQC and FSHQC
experiments were carried out on a cryoprobed Bruker Avance 500
Spectrometer at $T$ $=$ 300 $K$. Sample conditions were 150$\mu$M
Sso AcP in 15 mM phosphate buffer (pH 7.0, 7.3 and 7.6). The
low buffer concentration was to avoid sensitivity reduction in the
cryoprobe. Water suppression was by excitation sculpting
\cite{hdx2} or 3-9-19 watergate. Quadrature detection was by TPPI
\cite{hdx3}. Spectral widths were 7200 Hz in t$_{2}$ dimension and
2000 Hz in t$_{1}$. All spectra were composed of t$_{2}$ = 1024
$\times$ t$_{1}$ = 128 complex points. Data was processed using
the NMRPipe \cite{hdx4} and analyzed using Sparky \cite{hdx5}.

\textbf{CLEANEX-PM experiments.} Proton/proton exchange
measurements used the (CLEANEX-PM)-FHSQC pulse program introduced
by Mori et al. in 1998 \cite{hdx6}. CLEANEX mixing times were
between 10 and 50 ms. Assignment of peaks visible in
(CLEANEX-PM)-FHSQC spectra was by comparison with an assigned
reference FHSQC. CLEANEX experiments at pHs between 7.0 and 7.6
gave identical results. CLEANEX data were analyzed to give
quantitative exchange rates using the `initial slope' method
described by Hwang et al in 1998 \cite{hdx6}.

\textbf{FSHQC HDX experiments.} HDX measurements used
FSHQC pulse sequence introduced by Mori et al. in 1995 \cite{hdx7}. In all experiments, the instrument was pre-shimmed using a
blank sample (15 mM phosphate buffer in D$_{2}$O, pD 7.0, 7.3 and
7.6) in the NMR tube to be used in the HDX experiment. Sso AcP in
15 mM phosphate buffer (pH 7.0, 7.3 and 7.6) was concentrated to
$\sim$ 1.5 mM using Vivaspin centrifugal concentrators (MWCO
5,000) and quickly added to the `pre-shimmed', D$_{2}$O
solution-containing NMR tube and mixed by inversion. The interval
between mixing and the start of FHSQC acquisition was typically
around 1.5 min. This includes time for introduction of the sample
into the probe and brief manual re-shimming. Full FHSQCs were
collected every 7.5 min for the first 3 hours. After three hours,
the acquisition times were lengthened by increasing n$_{s}$ (the
number of scans/t$_{1}$). To acquire quantitative observed HDX rates k$_{obs}$,
time-dependent FHSQC peak intensities $I$ were normalized using
n$_{s}$ and fit to single exponential decay functions with
offsets:

\begin{equation}
I = ae^{-k_{obs}t} + b
\end{equation}

The extracted k$_{obs}$ are related to conformational flexibility
through primary sequence-specific extrinsic exchange rates
k$_{int}$, which were predicted using the Sphere software
\cite{hdx8}. When HDX occurs within the EX2 limit, the ratio
k$_{int}$/k$_{obs}$ (the `protection factor') is directly related
to conformational flexibility, expressed as an equilibrium between
an `open' (HDX-competant) and `closed' (HDX-incompetant) state for each backbone amide.

\section{Results and Discussion}

\subsection{Experimental HDX profile of Sso AcP}

\begin{figure}[h!]
\centering
{\includegraphics[width=.6\textwidth]{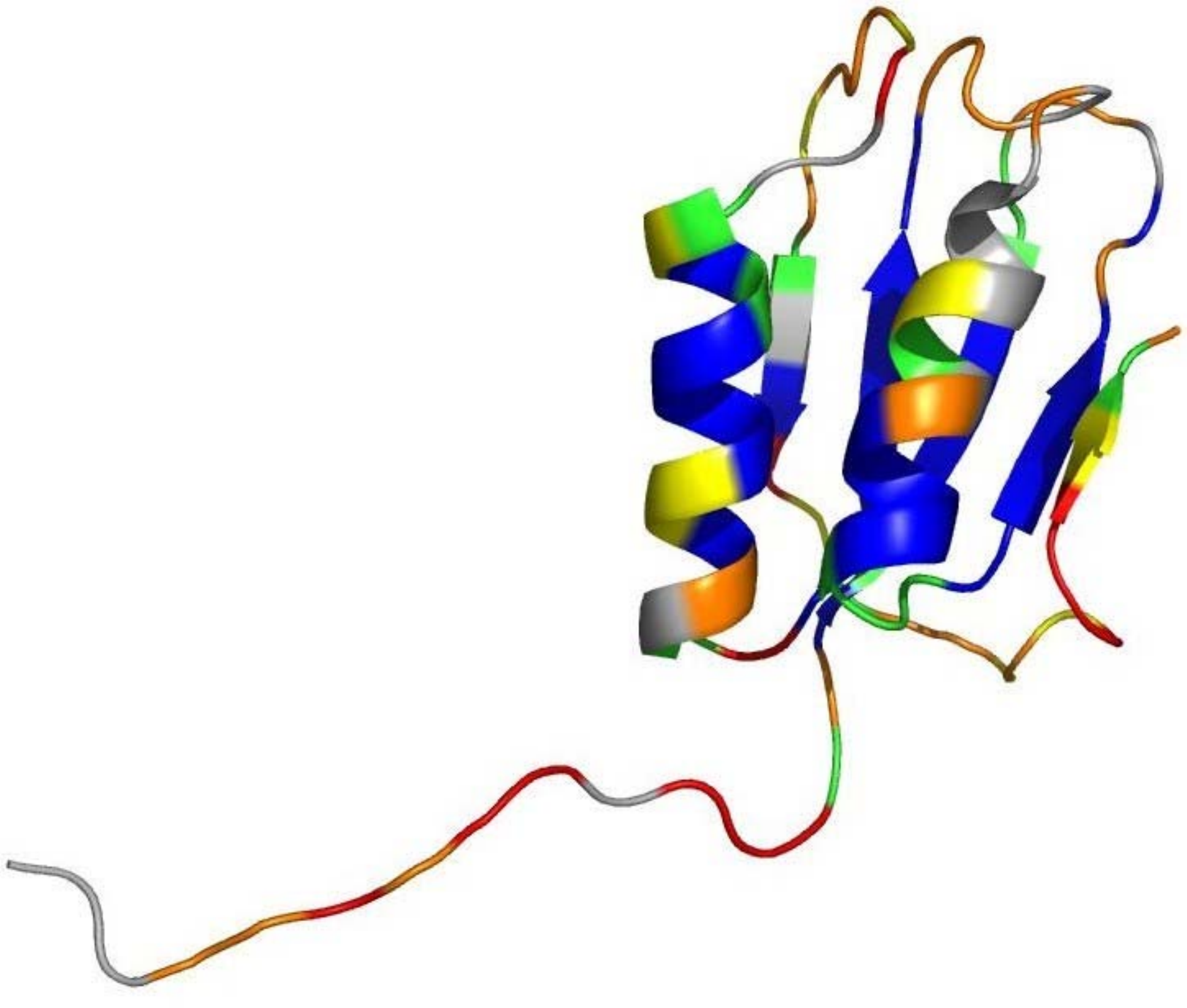}}
\quad
\caption{Hydrogen-Deuterium Exchange (HDX) profile of
Sso AcP mapped on to the
native structure (see text and supplementary information for details).} \label{fig:hdexchangeprofile}
    \end{figure}

In figure \ref{fig:hdexchangeprofile} the experimental HDX profile
of Sso AcP is overlaid on the 3-dimensional structure. We attained a high coverage native (ensemble) state HDX profile (87 \% of residues were covered) on this hyperthermophile protein at moderate temperatures. Red and
orange residues represent the very fast and fast exchangers,
respectively. These appear exclusively in the
loops and in the unstructured N-terminus tail region (first
12 residues), with the exception of one residue in the N-terminal end of $\alpha$-helix 2 and one in the centre of
$\alpha$-helix 1 (see figure \ref{fig:1y9o} for the labeling of secondary structures). Yellow residues represent the medium exchangers 
and blue the slowest exchangers (i.e. exchange not observed
over three weeks after exposure to D$_{2}$O). Residues colored green were detected in the FHSQC, but at insufficient signal-to-noise to provide reliable exchange data. Details of these results are provided in the supplementary information.

In the gray regions no HDX measurements could be made as no NMR signal was detected (or the residue is a proline); most of these residues are found in
unstructured parts of the protein. The notable
exception is the cluster of residues bounded by Gln25 and
Lys31, which are mostly found in the N-terminal end of
$\alpha$-helix 1. Of these only Val27 could be assigned in spite of the fact that these peaks were well dispersed in the reference spectrum.
This suggests that the N-terminal half of
$\alpha$-helix 1 is unstructured and flexible, and
likely undergoes large-amplitude `molten globule-like'
conformational dynamics. Residues undergoing
molten-globule-like conformational dynamics are frequently not
detectable by NMR due to extreme broadening of the HSQC signals.

Regions of the protein with high exchange rates (red and orange) correspond to the least protected regions, and regions with slowest exchange rates
(blue) are the most protected. 
Majority of the slowest exchangers are found in the
C-terminal half of $\alpha$-helix 1, $\alpha$-helix 2, and three central $\beta$-strands (strands 1, 2 and 3). $\alpha$-helix 2 has significantly more slow
exchangers than $\alpha$-helix 1. The two halves of $\alpha$-helix
1 are drastically different. Unlike the N-terminal half
which is very unstructured, the C-terminal half has only slow
exchangers and is very well protected from undergoing exchange.

\subsection{FIRST results on individual NMR models}

\begin{figure}[h!]
\centering
{\includegraphics[width=1\textwidth]{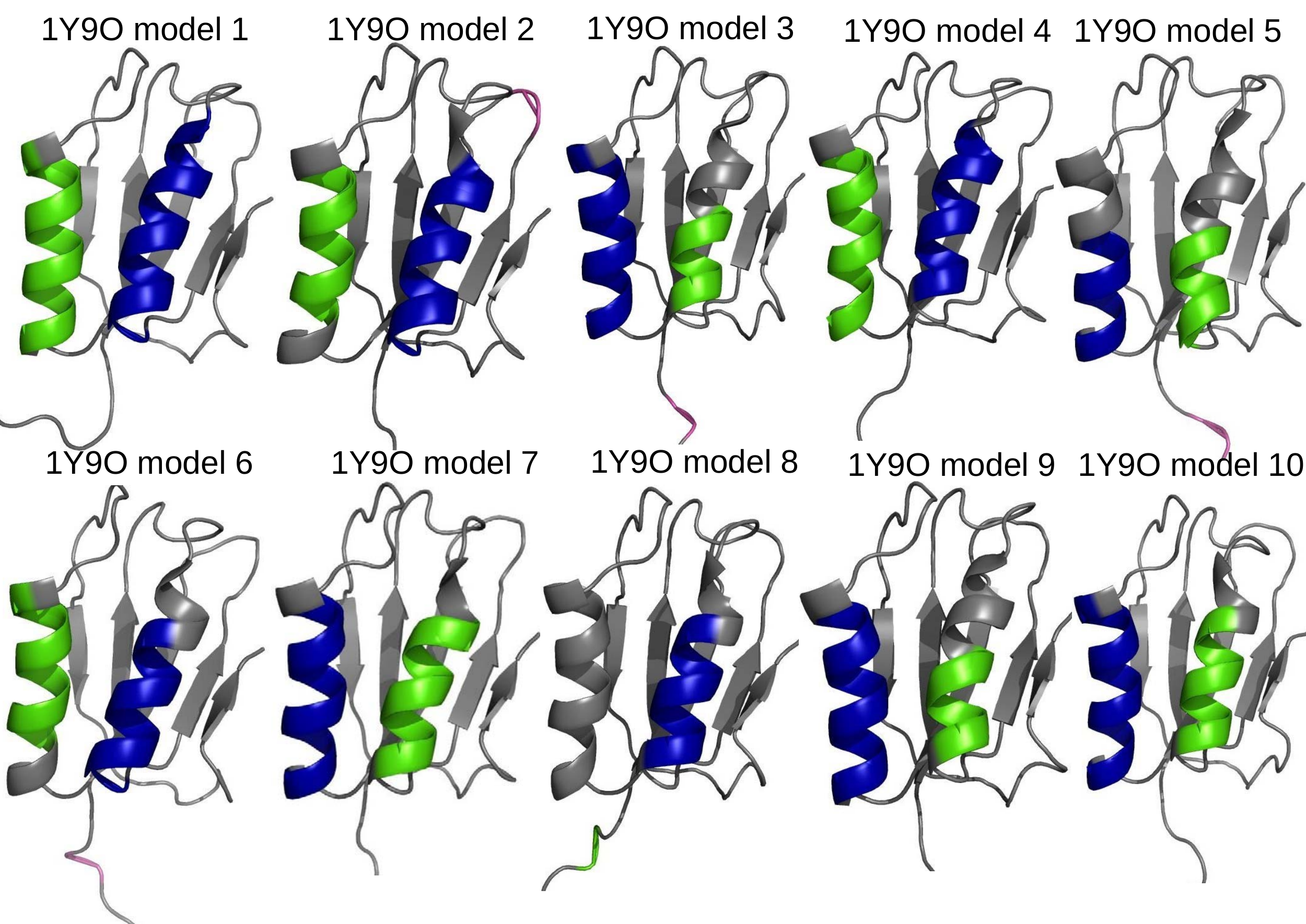}}
\quad
\caption{Rigid cluster decomposition using FIRST on NMR models of Sso Acp. The output of first 10 models from
the ensemble is displayed. The blue rigid cluster is the largest
rigid cluster, followed by the green rigid cluster. Note the
variation in the size of the rigid clusters among the NMR models.}
    \label{fig:1y9orcd10}
    \end{figure}

Before acquiring our FIRST-ensemble predictions on Sso AcP, we first obtained the regular `single snapshot' FIRST analysis on all 20 snapshots (models). The rigid cluster decomposition of the first ten models is displayed in figure
\ref{fig:1y9orcd10}.

There are only two large rigid clusters, 
corresponding to the blue and green regions in the two
$\alpha$-helices; flexible regions are couloured in gray. The $\beta$-sheet is flexible in all 20 models. In the most representative structure (model 1 -- typically the selected conformer in the previous FIRST studies on
NMR structures) both $\alpha$-helices are rigid.
Note there is a substantial difference in the size of these two rigid clusters over the ensemble.
The blue cluster is the largest rigid cluster, however there is a switch across the models in which $\alpha$-helix gets declared as
blue or green. In some models, parts of, or even an entire $\alpha$-helix
(model 8) are flexible. This shift is particularly evident in the
ends of $\alpha$-helices, notably in the N-terminal end of $\alpha$-helix
1, which has significantly weaker hydrogen bonding interactions
than its C-terminal end. 

It is known that small changes in conformation of atoms can lead
to altered hydrogen bonding geometry (hydrogen bond energy values)
(see \cite{Wells}) and a consequent change in the total number of
included hydrogen bond constraints. Hence, the variation in the size
of the rigid clusters across the different NMR models with slight structural heterogeneity is not surprising. The number of hydrogen bonds with energies less than $-$1 kcal/mol among the 20 models
of Sso AcP ranges from as few as 43 hydrogen bonds to as many as
54 hydrogen bonds. These differences will clearly have an effect
on the total number of remaining DOF from model to model, and on the size of the rigid clusters. These observations are similar in flavour to the study by Wells et al. \cite{Wells}, where the authors found that FIRST
analysis on structurally similar crystal
structures with the same
energy cutoff can produce different rigid cluster decompositions.

\subsection{FIRST-ensemble results}

\begin{figure}[h!]
\centering
{\includegraphics[width=.5\textwidth]{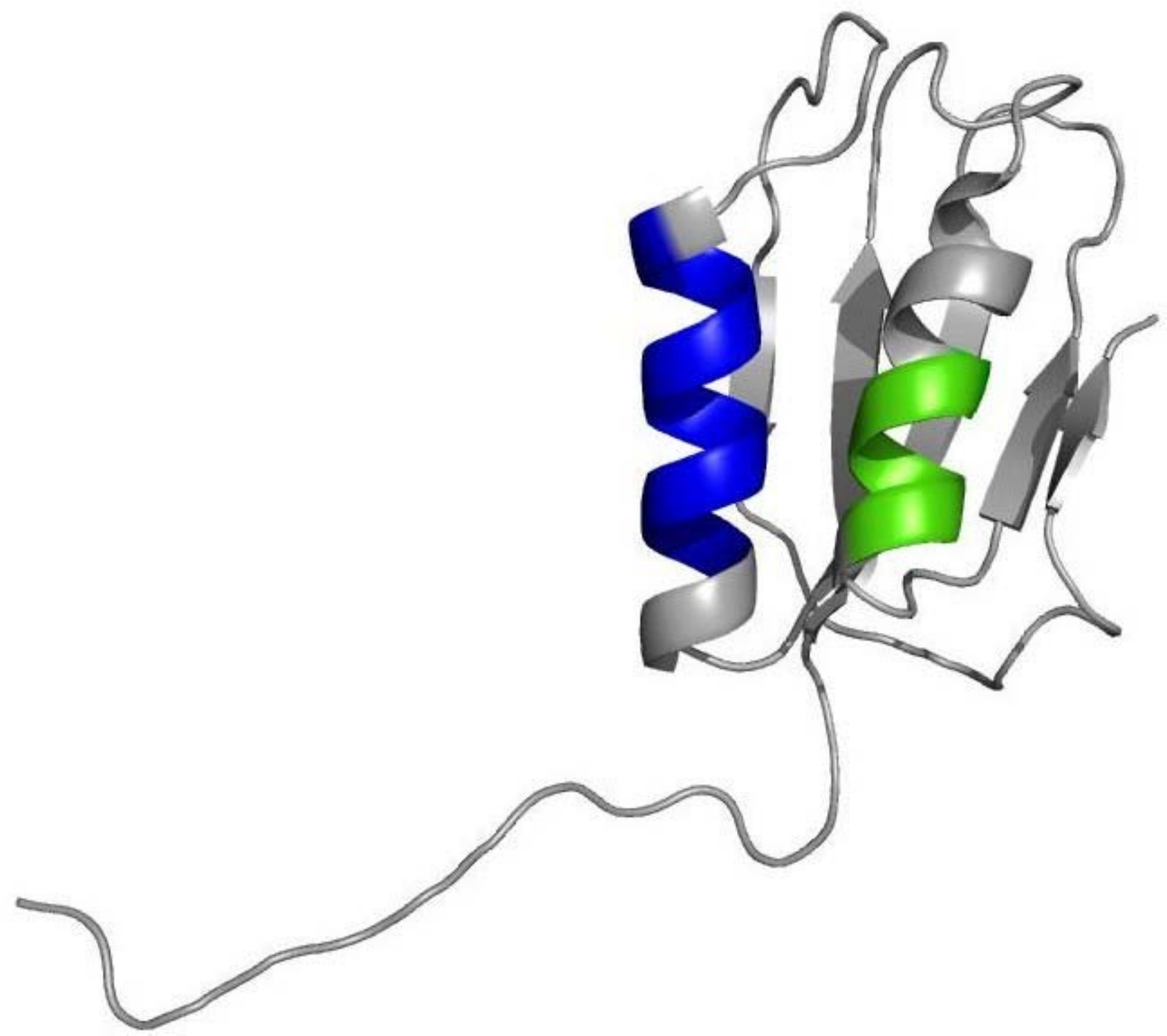}}
\quad
    \caption{Rigid cluster decomposition from FIRST-ensemble, Algorithm \ref{alg:firstensemble}. There are two main rigid clusters. Most of $\alpha$-helix 2 is rigid (blue), with the
    exception of its end points. On the other hand, $\alpha$-helix 1 is rigid on its C-terminal half (green), and flexible in the N-terminal half. The $\beta$-sheet is flexible.}
    \label{fig:1y9o_rigidityensemble}
    \end{figure}

The output of the FIRST-ensemble Algorithm \ref{alg:firstensemble}
on Sso AcP is provided in figure \ref{fig:1y9o_rigidityensemble}.
Over the ensemble, $\alpha$-helix 2 retains most of its
rigidity and becomes flexible only at its end points. On the other hand, roughly half of $\alpha$-helix 1 is rigid, corresponding to its C-terminal half. In the flexible N-terminal half of $\alpha$-helix 1 most of the
hydrogen bonds are very weak (i.e. their average strengths do
not pass the energy cutoff), or there is a lack of a sufficient number of persistent strong hydrogen bonds over the ensemble, which are modeled with less bars between hydrogen and acceptor atoms
(Algorithm \ref{alg:firstensemble} step 3 (b)). As expected, the
$\beta$-sheet remains flexible in the ensemble prediction.

\begin{figure}[h!]
\centering
{\includegraphics[width=.9\textwidth]{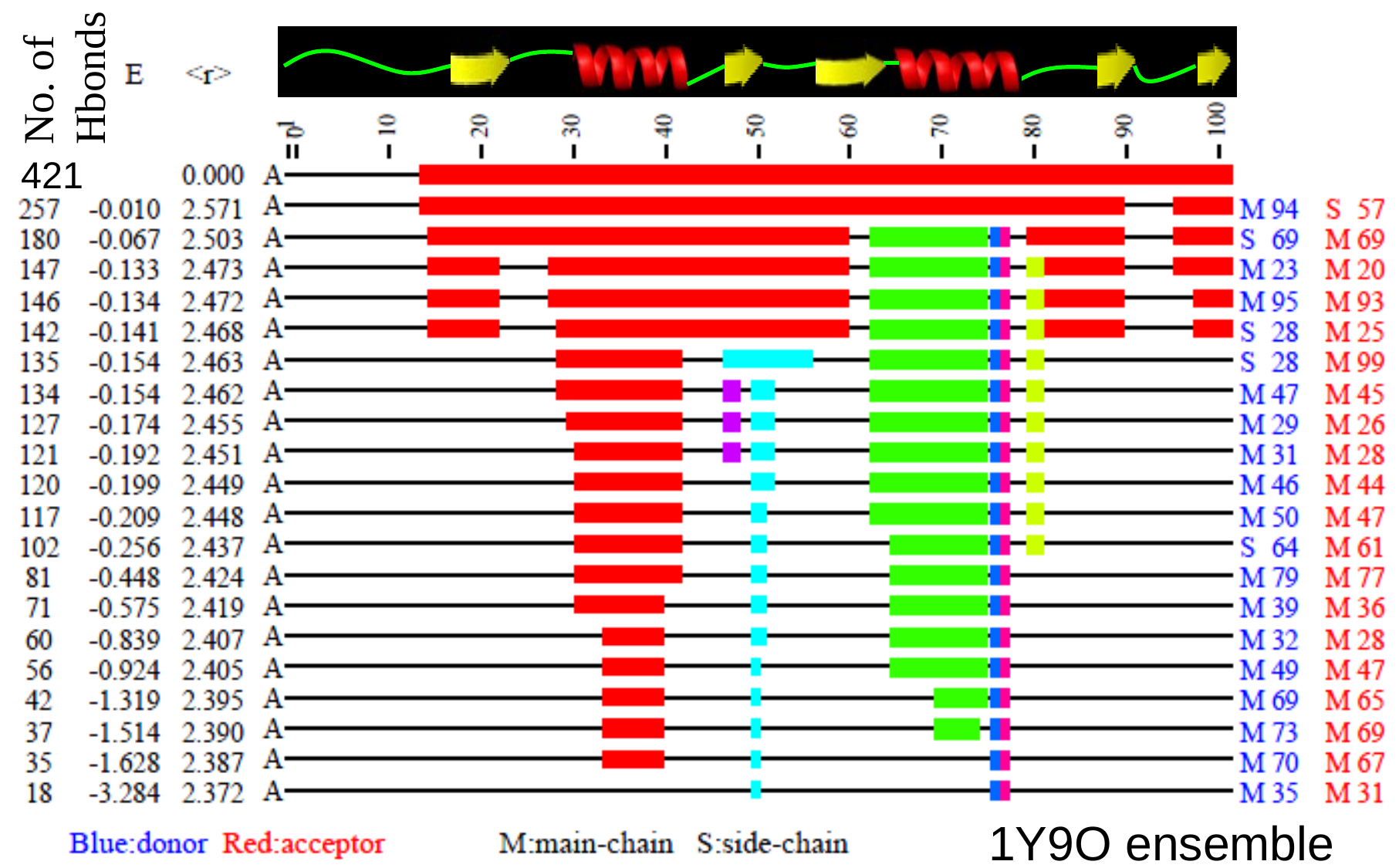}}
    \caption{Dilution plot of FIRST-ensemble (Algorithm \ref{alg:firstensemble}) on Sso AcP. Flexible
    regions are indicated with thin black lines, and rigid regions are indicated with blocks. The columns on the left-hand side are updated and display: the total number of remaining hydrogen bonds, the energy of the hydrogen bond that is currently broken in kcal/mol. The mean coordination number is also provided. The columns on the right represent the residue numbers of the donor and acceptor atoms of the broken hydrogen bond at each step. Initially, with inclusion of all potential hydrogen bonds from the entire ensemble, in the FIRST-ensemble prediction of Sso AcP the entire protein is rigid (top red block) with the exception of the long flexible tail at the N-terminus. Once most weak hydrogen bonds are broken (going down the dilution plot) several rigid clusters form, and it is evident from the dilution plot that $\beta$-sheets become flexible and essentially only two significant rigid clusters remain, corresponding to the two $\alpha$-helices.}
    \label{fig:averagedilutionplot}
    \end{figure}

Changes in the rigidity can be monitored by a gradual removal of
hydrogen bonds one by one (i.e. by lowering of hydrogen bond energy cutoff) in the order of strength, keeping all covalent and
hydrophobic interactions intact, and then redoing the rigidity analysis at each step identifying rigid and flexible regions. The change in rigidity can be visualized nicely using the hydrogen bond `dilution plot' \cite{flexgraph}. The dilution plot for FIRST-ensemble prediction on Sso AcP is
shown in figure \ref{fig:averagedilutionplot} 
(see \cite{cores, rigiditylost, flexgraph, Wells} for detailed explanation of dilution plots).

The fact that the $\beta$-sheet in Sso AcP is flexible and the
$\alpha$-helices are rigid is
concordant with the study by Whiteley \cite{countingout}, which
provides a precise rigidity-based characterization of
flexibility and the minimum number of constraints needed for simple
secondary structure motifs (i.e. rings, loops, $\alpha$-helices,
$\beta$-sheets and $\beta$-barrels) to become rigid in FIRST. With the exception of $\beta$-barrels, this study reminds us that an isolated $\beta$-sheet will only become rigid when there are a large number of strands or fewer strands but with longer lengths. On
the other hand, an isolated $\alpha$-helix (or part of) will attain its
rigidity much more easily than a $\beta$-sheet \cite{countingout}. Our rigidity analysis of Sso AcP is also consistent
with previous FIRST analysis on other proteins. For instance,
Wells et al. \cite{Wells} have observed that as weak hydrogen
bonds are broken (diluted), $\alpha$-helices retain their rigidity
much longer than $\beta$-sheets.

\begin{figure}[h!]
\centering
{\includegraphics[width=.3\textwidth]{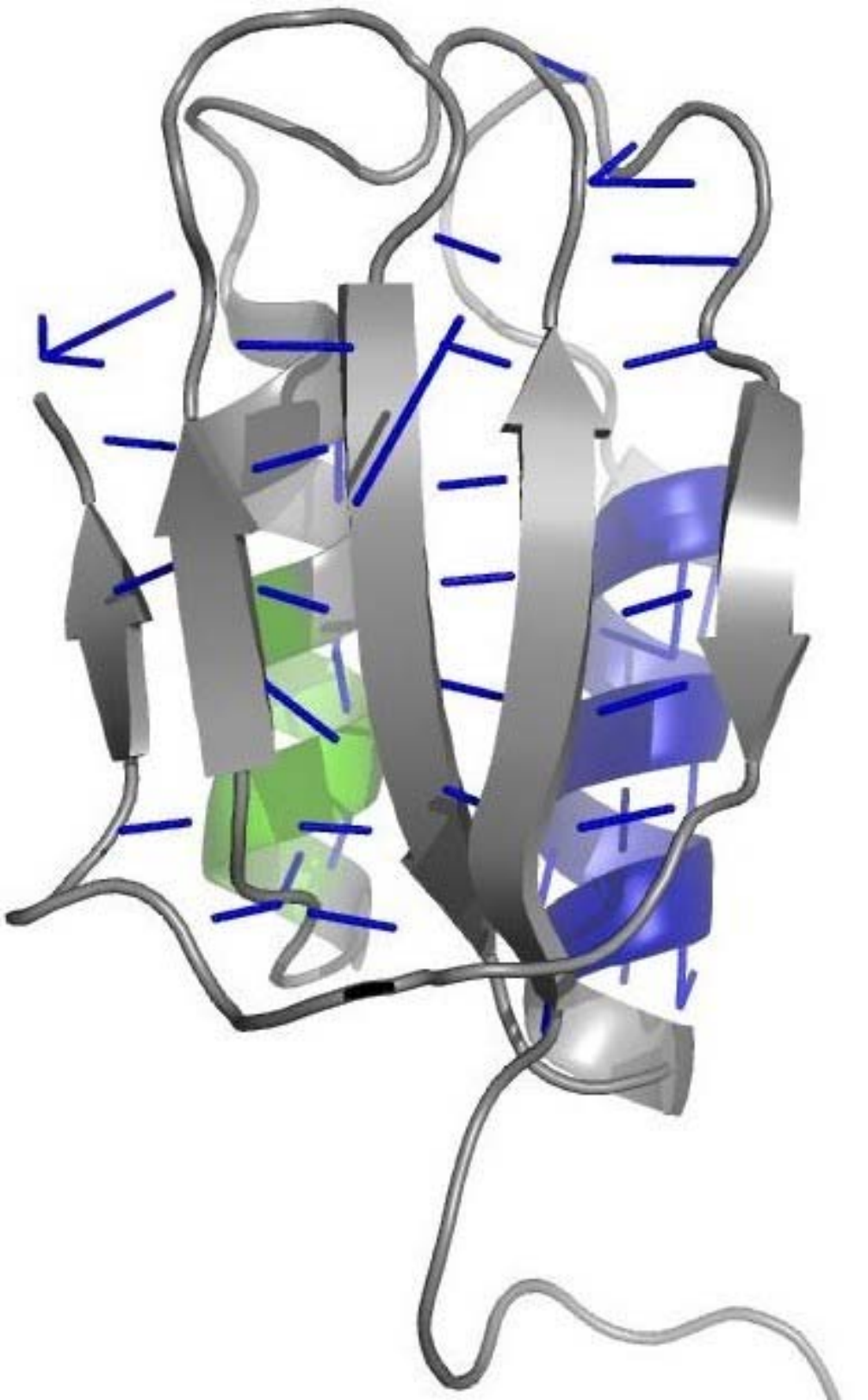}}
\quad
    \caption{Output of FIRST-ensemble, same as in figure~\ref{fig:1y9o_rigidityensemble} with a different orientation. There are a significant number of
    strong inter-strand hydrogen bonds (indicated with blue lines) that are present in the flexible $\beta$-sheet.}
    \label{fig:1y9o_rigidityensemble_hbonds}
    \end{figure}

Even though the $\beta$-sheet of Sso AcP is flexible, it has a
well-defined structure and a high number of persistent strong
inter-strand hydrogen bonds over the entire ensemble. In figure \ref{fig:1y9o_rigidityensemble_hbonds} we have
displayed the hydrogen bond constraints that are
included in the FIRST-ensemble analysis. The strength of these persistent inter-strand hydrogen bonds in
the $\beta$-sheet is also supported by the anti-parallel arrangement of the
four $\beta$-strands 1 to 4, which gives rise to the preferred nearly planar hydrogen bond geometry between the NH and CO groups.

Normally $\beta$-sheets are rigidified with hydrophobic interactions and side chain hydrogen bonds. This occurs when $\beta$-sheets are
well-packed against other $\beta$-sheets or $\alpha$-helices, as is for instance
the case in the amyloid-fibril formation via stacking of
$\beta$-sheets, where there is both strong inter-strand and
inter-$\beta$-sheet interactions \cite{amyloid}. The main reason the $\beta$-sheet of Sso AcP is flexible is due to the lack of these additional constraints within the strands or between the $\beta$-sheet and $\alpha$-helices. Most of these hydrogen bonds are very weak with energies close to 0 kcal/mol. It seems that the importance of the side chain hydrogen bonding interactions and hydrophobic contacts to rigidity of $\beta$-sheets has not been explicitly commented or investigated in the FIRST literature.

The clear advantage of the FIRST-ensemble prediction over the traditional FIRST runs performed on NMR files using a single snapshot is that
FIRST-ensemble incorporates the structural information from the entire NMR ensemble. By averaging the hydrogen bonding strengths and incorporating persistence of hydrogen bonds from all structures, with FIRST-ensemble prediction we can capture the variations in the rigid clusters between individual models of Sso AcP; (Note that averaging hydrophobic interactions is not essential for ensemble rigidity predictions - see Supplementary information). Furthermore, in contrast to the previous FIRST study \cite{gohlkekuhn}, where the authors performed rigidity analysis on all individual MD generated snapshots, and then averaged rigidity results (i.e. flexibility index), our FIRST-ensemble prediction is attained using only a single adapted FIRST run. This way we can continue to utilize and take advantage of the fast computational speed of FIRST, and at the same time incorporate the structural information from entire ensemble.

As was discussed earlier, rigidity and flexibility prediction on its own does not lead to a conclusive measure of HDX as we also need solvent accessibility. However, it is already evident that the FIRST-ensemble predictions are much better matched with the experimental HDX data on Sso AcP than the traditional FIRST run using a single NMR snapshot. In the FIRST prediction on the most representative model, both $\alpha$-helices are entirely rigid, while in the FIRST-ensemble prediction, $\alpha$-helix 2 is mostly rigid with its ends becoming flexible,
and only half of $\alpha$-helix 1 comes up as rigid. This is in agreement with the experimental HDX data, where roughly half of $\alpha$-helix 1 and
most of $\alpha$-helix 2 (except its ends) are protected from HDX (Figure \ref{fig:hdexchangeprofile}). It is
certainly not uncommon for parts of $\alpha$-helices to be unstructured
and flexible, but considering the hyporthermophile nature of this
protein, it is remarkable that FIRST-ensemble is able to capture
sub-structural flexibility of $\alpha$-helix 1. Clearly, in using only one
structure out of the ensemble as an input into FIRST, important
structural variations across the ensemble are lost, particularly the
conformational variations in the N-terminal half of $\alpha$-helix 1. On
both $\alpha$-helices, the FIRST-ensemble predicted rigid regions
give a very good indication of the regions that will be protected
from HDX. Since the rest of the protein is flexible, we cannot yet draw any definite comparisons or conclusions with respect to the HDX data (see solvent accessibility of Sso AcP).





\subsection{Solvent Accessibility ensemble data}

\begin{figure}[h!]
\centering
{\includegraphics[width=1\textwidth]{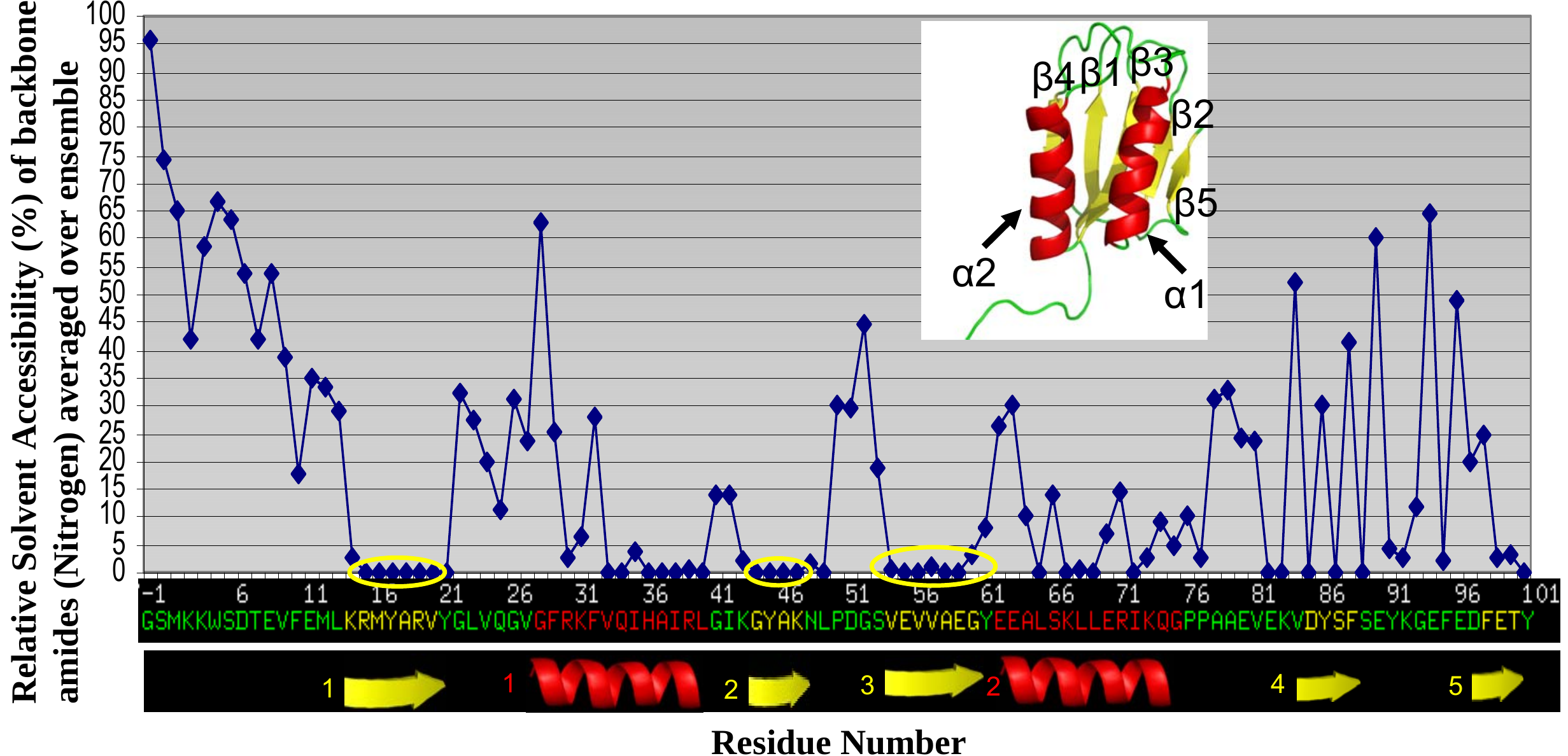}}
\quad
    \caption{Solvent accessibility (RSA) of Sso AcP of backbone amide (NH) averaged over the ensemble; obtained from
WHATIF \cite{whatif1}. Note the three central $\beta$-strands (strand 1, 2, and
    3, circled in yellow) are almost completely buried.}
        \label{fig:accessibilityplot}
    \end{figure}

\begin{figure}[h!]
\centering
   \subfigure[] {\includegraphics[width=.5\textwidth]{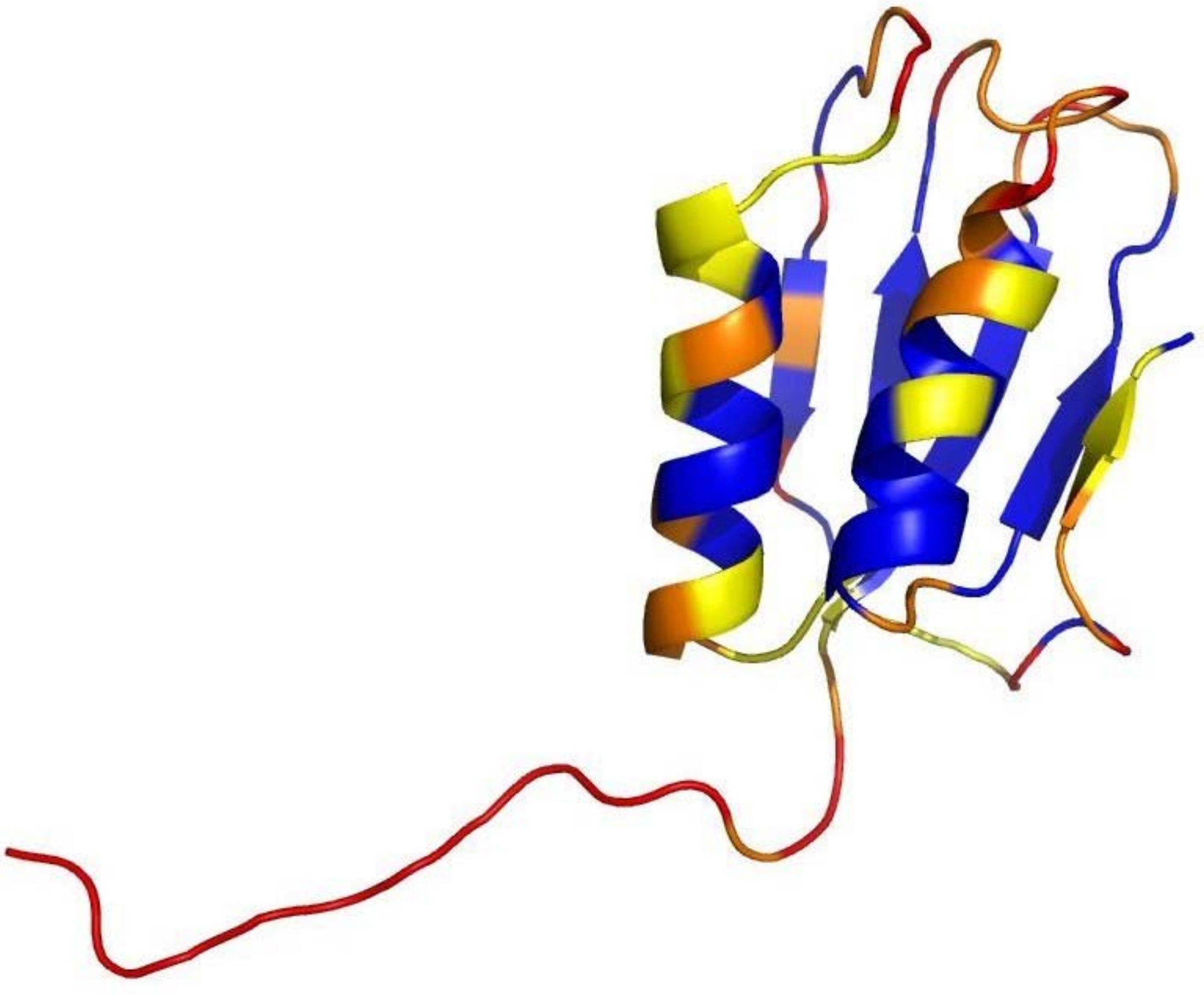}} \quad
      \subfigure[] {\includegraphics[width=.24\textwidth]{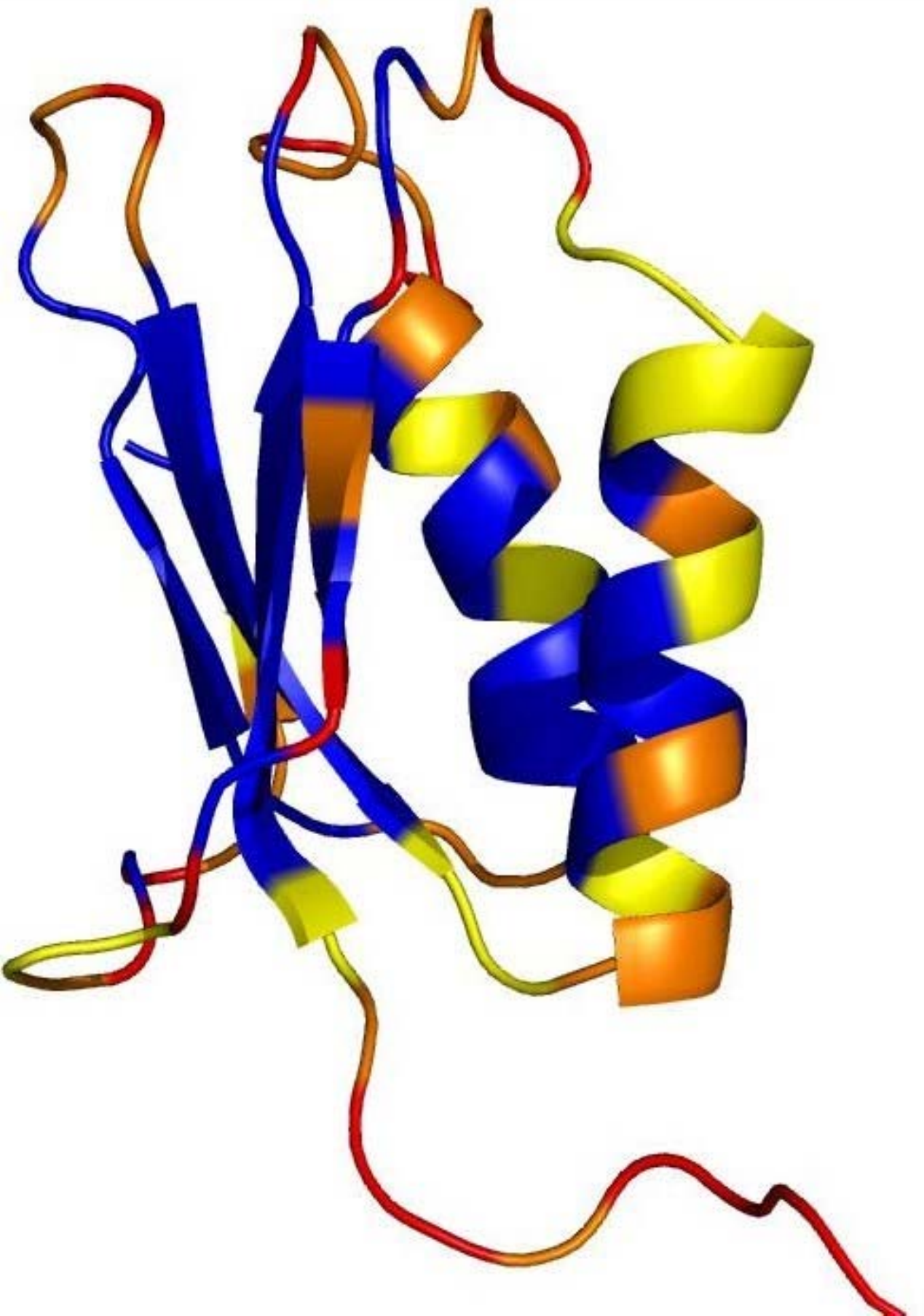}} \quad
    \caption{Solvent accessibility (RSA) of Sso AcP of backbone amide (NH) averaged over the ensemble. The blue regions are almost completely buried (solvent inaccessible)
    (RSA $<=$ 4\%), followed by yellow regions - somewhat exposed, orange - mostly exposed, and red - almost completely
    exposed (a). Same display with different orientation (b) which indicates the mixed accessibility of $\alpha$-helices, with the more buried side pointing towards the
    interior of the protein.}
    \label{fig:accessibility4colours}
    \end{figure}

The plot of solvent accessibility (RSA) for the backbone amide
Nitrogens of Sso Acp, averaged over the ensemble, is given in
figure \ref{fig:accessibilityplot}. In figure \ref{fig:accessibility4colours} we have coloured the structure with the solvent accessibility
colouring scheme as described in section \ref{sec:accessibility}. The first striking observation is that the three $\beta$-strands
(strands 1, 2 and 3) are completely buried. In fact, almost every residue in these three strands has 0\% average solvent accessibility. Having no backbone amide (NH) accessibility over the ensemble suggests that these strands should
be well protected from undergoing HDX. Indeed, the experimental
HDX data confirms that the three central $\beta$-strands (three blue strands in figure \ref{fig:hdexchangeprofile}) are the slowest exchangers (best protected from undergoing HD-exchange). The two side strands are more solvent accessible. $\beta$-strand 4, located behind $\alpha$-helix 2 has both buried and almost completely exposed amides. At the C-terminus of strand 4, the backbone amide Nitrogen of residue Phe87 is almost completely exposed (47 \%
accessibility) as it is directed towards the solvent. It is bound
by two completely buried (0 \% accessible) Ser86 and Ser88,
whose amide protons are pointing towards the Sso AcP structural
core. This is also in good correspondence with the experimental HDX,
where a `Very Fast' Phe87 exchanger is bounded by two
`Very Slow' serine exchangers (Figure
\ref{fig:hdexchangeprofile}). $\beta$-strand 5, the short strand
located at the opposite edge of the $\beta$-sheet appears to be
overall more solvent accessible. This again is in reasonable
agreement with the experimental HDX.

We had indicated earlier that we will not use solvent accessibility data on $\alpha$-helices, as it is not a reliable
indicator of HDX due to the amphipathic nature of
$\alpha$-helices. As anticipated, we found that the $\alpha$-helices of Sso AcP have
mixed solvent accessibility. This becomes visually apparent with the side view orientation of the protein (Figure
\ref{fig:accessibility4colours} (b)). Here we observe that the two
sides of $\alpha$-helices that are facing one another or the sides facing the $\beta$-sheet are mostly buried, while the opposite sides are more solvent exposed. In the N-terminal half of $\alpha$-helix 1, several residues have very low accessibility (Arg30 and Val33 are almost completely buried,
Ly31 is only somewhat exposed), yet we have already seen that
this flexible half of $\alpha$-helix 1 is very unstructured, and from
experimental HDX evidence it does not represent a slow exchanging
part of the protein. Similarly, solvent accessibility data on
$\alpha$-helix 2 would not be a good predictor of HDX, and is not well
agreed with the experimental HDX profile.

In addition to the amphipathic nature of $\alpha$-helices, we can envision
other reasons why solvent accessibility on $\alpha$-helices is
not a sufficient measure to probe HDX. Unlike $\beta$-sheets,
which can be predicted as flexible by FIRST and yet have a
well-defined structure with strong inter-strand hydrogen bonding
network, substructural flexibility in an $\alpha$-helical
structure is a direct consequence of weak $\alpha$-helical hydrogen bonding interactions. In the flexible regions of an $\alpha$-helix, the backbone amides are not involved in hydrogen bonding for the maintenance of secondary structure. So, even if part of the flexible $\alpha$-helix is computationally predicted to be buried on a snapshot(s), this region of a $\alpha$-helix can still potentially undergo backbone motions and turn a
previously buried backbone amide proton into a solvent exposed
amide proton. In such a case HDX could still occur.

In summary, solvent accessibility on Sso AcP gives a good estimate
of HDX on the $\beta$-sheet, and a poorer prediction on
$\alpha$-helices as was expected. The remaining flexible
regions (loops and the long flexible tail) are mostly solvent accessible, and still correspond well with the experimentally determined fast HD-exchangers. It is well known that flexible and highly
mobile regions in the protein are likely to be solvent accessible
\cite{residueflexibilitysolventaccessibility}. As we are
calculating the average solvent accessibility over the ensemble, any highly flexible region that retains its solvent
accessibility over the entire ensemble has a higher likelihood to
remain solvent accessible. 

\subsection{Combined rigidity and solvent accessibility predictions of HDX on Sso AcP}

Rigidity (FIRST-ensemble) predictions and solvent accessibility measures are valuable tools to computationally probe HDX. We have seen that on their own, rigidity or solvent accessibility, would lead to an inefficient overall predictor. By incorporating both analysis, an improved prediction of HDX can be achieved.

\begin{figure}[h!]
\centering
   \subfigure[] {\includegraphics[width=.47\textwidth]{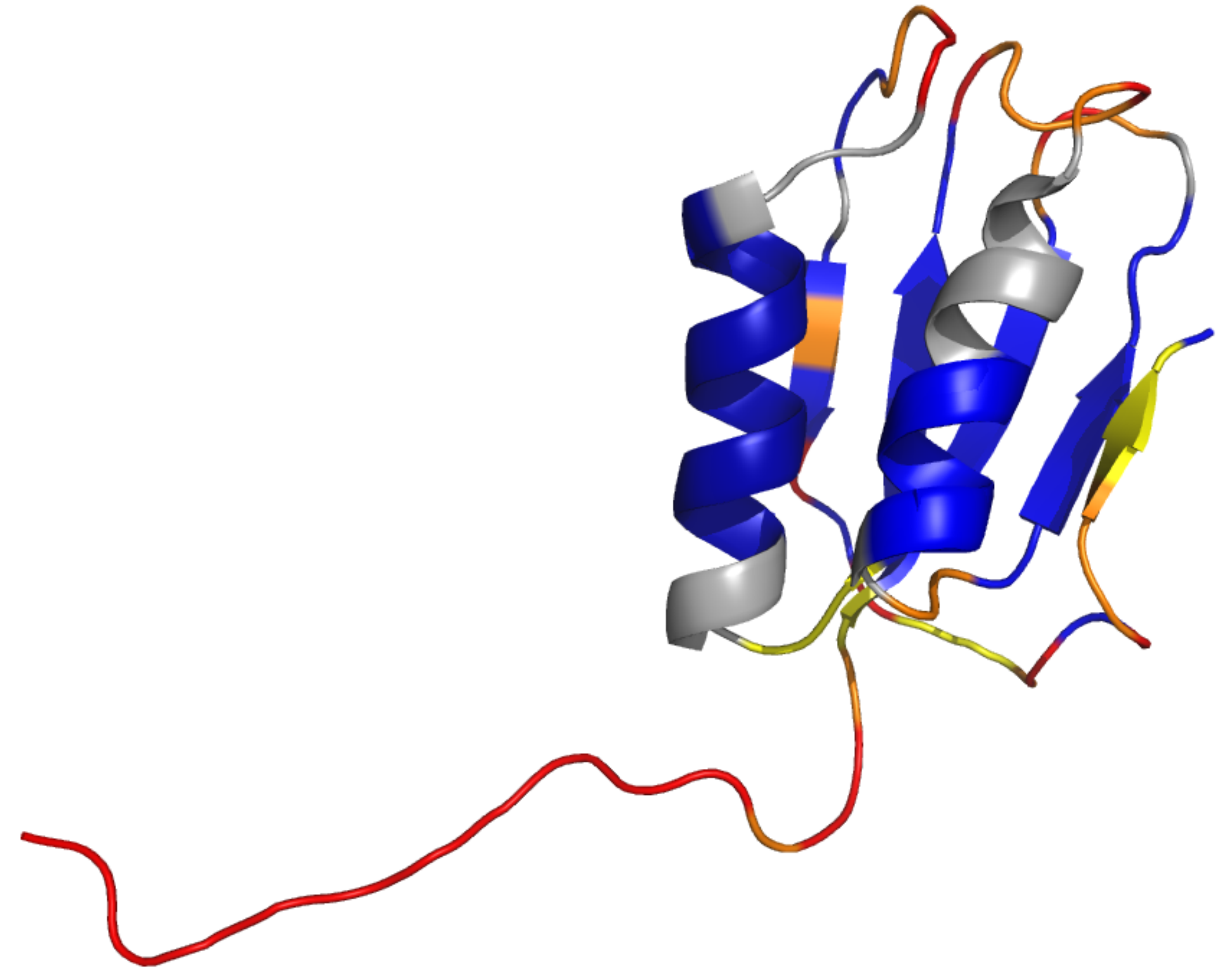}} \quad
   \subfigure[] {\includegraphics[width=.47\textwidth]{figure4.pdf}} \quad
    \caption{Combined rigidity (FIRST-ensemble) and solvent accessibility on Sso AcP, see Algorithm \ref{alg:rigidityaccessibilitycombined} (a), and
    experimentally derived HDX profile of Sso AcP (b). The blue regions (computationally predicted to not undergo HDX) in (a) (obtained by combining the
    rigid clusters and buried residues) match exceptionally well with the experimentally determined slow
    exchangers. The remaining regions are also in a good agreement.}
    \label{fig:rigidityaccessibility4colours1}
    \end{figure}

\begin{figure}[h!]
\centering
{\includegraphics[width=1\textwidth]{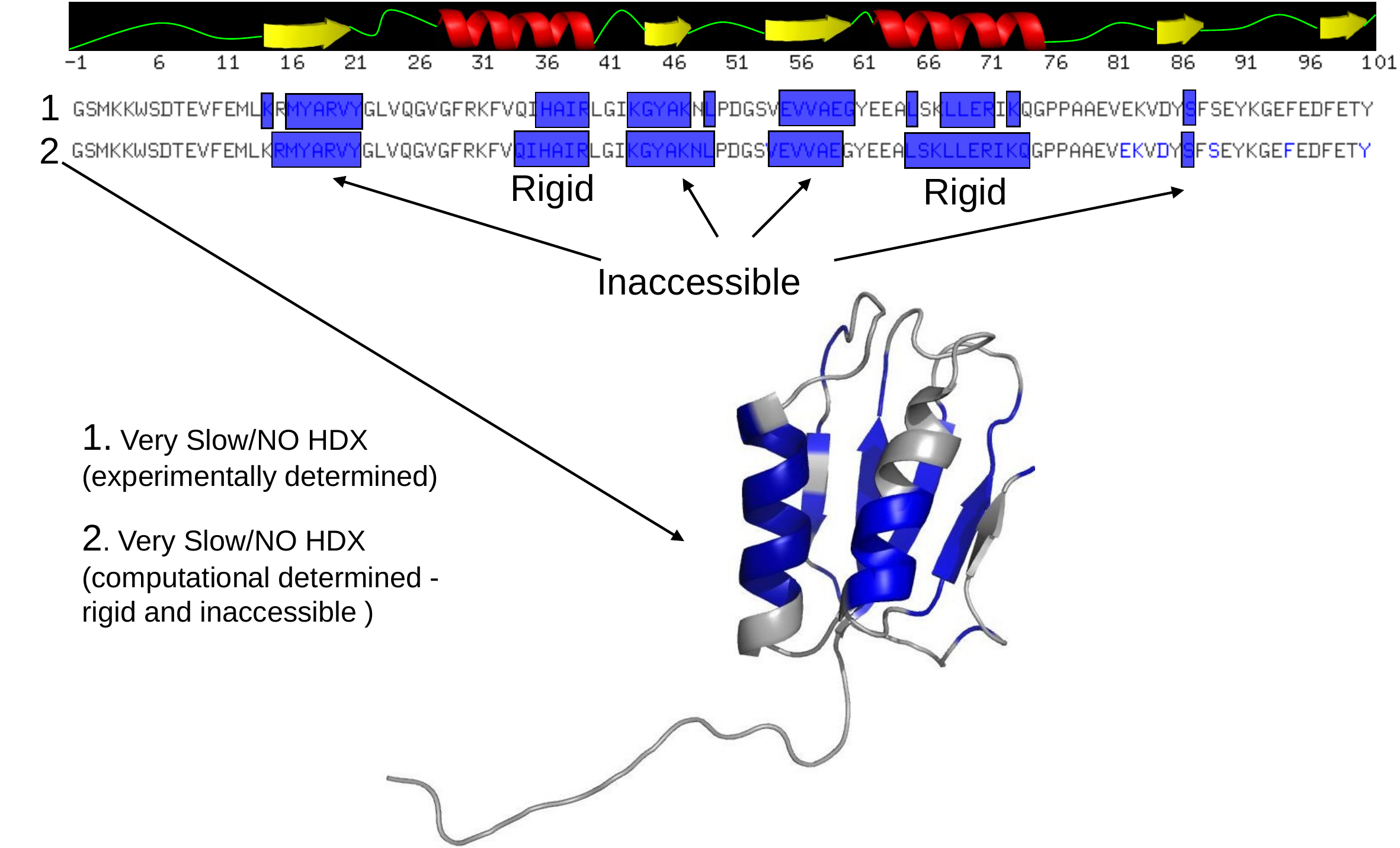}}
\quad
\caption{Comparison of the computationally predicted secondary structure residues that are
least likely to undergo exchange with the experimentally
determined slowest
exchangers is illustrated on the backbone protein chain.
In line 1 the blue regions represent the experimentally determined
very slow exchangers (most protected residues). Line 2 depicts the computationally predicted slowest exchangers composed of rigid and almost
completely buried residue. The computational predictions are remarkably well matched with experimentally determined very slow exchangers on all the major secondary structure regions.}
    \label{fig:rigid_inaccessible_hdsequencemap}
    \end{figure}

The combined FIRST-ensemble rigid clusters and solvent
accessibility on Sso AcP is shown in figure
\ref{fig:rigidityaccessibility4colours1} (a) (see Algorithm
\ref{alg:rigidityaccessibilitycombined}). We recall that
the blue region in the $\alpha$-helices corresponds to the two
rigid clusters found by FIRST-ensemble algorithm (Figure
\ref{fig:1y9o_rigidityensemble}). The remaining blue regions (notably the three central $\beta$-strands) and some residues in
$\beta$-strand 1 represent the `solvent inaccessible' regions.
These (combined) blue regions are the predicted regions in the
protein that will be protected from undergoing HDX (i.e. slowest exchangers). Comparing this computational prediction with the experimentally obtained HDX profile of Sso AcP, the predicted HDX protected regions
are in very good agreement with the experimentally determined very slow exchangers. For richer visual display of these results, in figure \ref{fig:rigid_inaccessible_hdsequencemap} we have presented both computational predictions and experimentally determined slow exchangers on a 1-dimensional (backbone) representation of the protein.

In summary, our computational approach gives a very good
prediction of the regions that remain most protected from
exchange which is supported by experimental HDX data. We have
obtained a good prediction on the loops and the
N-terminus tail, but as expected these regions are highly flexible
and unstructured (a few stable hydrogen bonds), and are
harder to probe than the secondary structures. Our findings on Sso
AcP support our hypothesis that combined rigidity-solvent accessibility predictions can be used to probe HDX.

\section{Conclusion and Outlook}

The aim of this study was twofold: 
to introduce a novel technique for computing rigidity of
ensembles using the FIRST method and to show that combining rigidity
and solvent accessibility can lead to good fast computational
predictions of HDX, as it was
applied on a hyperthermophile protein Sso AcP. 

We found that there is a significant variation in the rigid cluster decompositions
across individual NMR models. FIRST-ensemble prediction
incorporates the structural information and variations from all
the models and we have shown that the ensemble rigid cluster
decomposition is best matched with experimental HDX on Sso AcP.
From our findings on Sso AcP, and general analysis on FIRST in
\cite{Wells}, we suggest that the FIRST rigidity analysis on NMR
ensembles or other sources of ensemble data (i.e. MD snapshots) should be based on the FIRST-ensemble predictions to make them more robust. As part of future work, for comparison and further validation, it would be valuable to apply our
FIRST-ensemble algorithm on a larger class of NMR proteins, where
it would be desired to have comparable experimental data, such as the high
coverage native state HDX profile on Sso AcP we obtained in this study. The methods and techniques developed in this paper should further enhance the capability of FIRST and offer new tools and research avenues in predicting rigidity of ensembles. In other current work \cite{peptidecoarsegraining},
using rigidity adapted coarse graining MD simulations, initial results suggest that FIRST-ensemble analysis on NMR file of hemagglutinin fusion-peptide gives an improved rigidity prediction compared to the single snapshot FIRST analysis.

Both rigidity and solvent accessibility are important tools in
probing HDX, but best prediction is achieved when these
measures are combined.  We have shown
that our combined ensemble rigidity/accessibility algorithmic
predictions on Sso AcP is well matched with the experimental HDX
profile of Sso AcP. In the last 10 years there has been a rapidly growing interest in
applying rigidity based techniques to study both flexibility and
motions of proteins, RNA and DNA. To the best of our knowledge, this is the first study that incorporates rigidity and accessibility as a computational method
for predicting HDX. The clear advantage of these techniques is that they offer very fast computational technique in probing an expensive and laborious experimental method.

\medskip
\noindent {\bf Acknowledgements.}

We thank Walter Whiteley for many extended discussions on applications of rigidity theory to protein flexibility and valuable feedback on this paper. We are also grateful to Fabricio Chiti for supplying the Sso AcP expression vector.





\newpage
\newpage

\section*{References}
\end{document}